\documentclass[10pt]{article}
\pdfoutput=1
\usepackage{geometry}
\usepackage{amssymb}
\usepackage{textcomp}
\usepackage{amsmath}
\usepackage{bm}
\usepackage{hyperref}
\usepackage{graphicx}
\usepackage{cancel}
\usepackage{times}
\usepackage{epsfig}
\usepackage{xcolor}
\usepackage{colortbl}
\usepackage{slashed}
\usepackage{booktabs}
\usepackage{extarrows}
\usepackage{multirow}
\usepackage{latexsym}
\definecolor{mygray}{gray}{.95}
\usepackage[title]{appendix}
\allowdisplaybreaks[4]

\newcommand{\TeV}{{\rm TeV}}
\newcommand{\GeV}{{\rm GeV}}

\newcommand{\LEW}{\Lambda_{\rm EW}}
\newcommand{\LNP}{\Lambda_{\rm NP}}
\newcommand{\T}{{\rm T}}

\newcommand{\EM}{{\rm EM}}

\newcommand{\calA}{{\cal A}}
\newcommand{\calL}{{\cal L}}
\newcommand{\calO}{{\cal O}}
\newcommand{\calU}{{\cal U}}
\newcommand{\calV}{{\cal V}}

\begin{document}
\baselineskip=16pt

\pagenumbering{arabic}

\vspace{1.0cm}

\begin{center}
{\Large\sf Extending low energy effective field theory with a complete set of dimension-7 operators }
\\[10pt]
\vspace{.5 cm}

{Yi Liao~$^{a,c}$\footnote{liaoy@nankai.edu.cn},~
Xiao-Dong Ma~$^{b}$\footnote{maxid@mail.nankai.edu.cn},~
Quan-Yu Wang~$^{a}$\footnote{2120170120@mail.nankai.edu.cn},
}

{$^a$~School of Physics, Nankai University, Tianjin 300071, China
\\
$^b$ Department of Physics, National Taiwan University, Taipei 10617, Taiwan\\
$^c$ Center for High Energy Physics, Peking University, Beijing 100871, China}

\vspace{2.0ex}

{\bf Abstract}
\end{center}

We present a complete and independent set of dimension-7 operators in the low energy effective field theory (LEFT) where the dynamical degrees of freedom are the standard model five quarks and all of the neutral and charged leptons. All operators are non-Hermitian and are classified according to their baryon ($\Delta B$) and lepton ($\Delta L$) numbers violated. Including Hermitian-conjugated operators, there are in total $3168$, $750$, $588$, $712$ operators with $(\Delta B,\Delta L)=(0,0),~(0,\pm 2),~(\pm 1,\mp 1),~(\pm 1,\pm 1)$ respectively. We perform the tree-level matching with the standard model effective field theory (SMEFT) up to dimension-7 (dim-7) operators in both LEFT and SMEFT. As a phenomenological application we study the effective neutrino-photon interactions due to dim-7 lepton number violating operators that are induced and much enhanced at one loop from dim-6 operators that in turn are matched from dim-7 SMEFT operators. We compare various neutrino-photon scattering cross sections with their counterparts in the standard model and highlight the new features. Finally, we illustrate how these effective interactions could arise from ultraviolet completion.

%%%%%%%%%%%%%%%%%%%%%%%
\section{Introduction}
%%%%%%%%%%%%%%%%%%%%%%%

While neutrino mass and dark matter provide evidence for physics beyond the standard model (SM), persistent searches for new heavy particle production have hitherto yielded a null result. In this circumstance, effective field theory (EFT) offers an appropriate and universal approach to quantifying unknown effects of possibly very heavy new particles on the interactions of SM particles at relatively low energies. In this framework, i.e., the standard model effective field theory (SMEFT), the standard model appears as the leading interactions that are generally augmented by an infinite tower of effective interactions that involve higher and higher dimensional operators and are more and more suppressed by heavy particles masses. The precise measurements and severe constraints on these effective interactions will shed light on possible form of new physics.

Suppose that certain new physics scale $\LNP$ is significantly higher than the electroweak scale $\LEW\sim 10^2~\GeV$ and that there are no particles other than the SM ones of a mass around or below $\LEW$. The effective field theory between the scales $\LNP$ and $\LEW$ is then the SMEFT that includes all SM fields and satisfies the complete gauge symmetry $SU(3)_C\times SU(2)_L\times U(1)_Y$. Since it is an EFT at low energy compared to $\LNP$, it can be organized by the dimensions of operators involved in effective interactions. The bases of complete and independent operators have now been known at dimension 5 (dim-5)~\cite{Weinberg:1979sa}, dimension 6~\cite{Buchmuller:1985jz,Grzadkowski:2010es},  dimension 7~\cite{Lehman:2014jma,Liao:2016hru}, and dimension 8~\cite{Lehman:2015coa, Henning:2015alf, Li:2020gnx,Murphy:2020rsh}, and the one-loop renormalization of those basis operators due to the SM interactions has been accomplished up to dimension 7 in Refs.~\cite{Babu:1993qv,Antusch:2001ck, Grojean:2013kd,Elias-Miro:2013gya,Elias-Miro:2013mua,
Jenkins:2013zja,Jenkins:2013wua,Alonso:2013hga,Alonso:2014zka,
Liao:2016hru,Liao:2019tep}. As the dimension of operators goes up further, the number of basis operators increases horribly fast~\cite{Henning:2015alf}; for recent efforts on basis operators of even higher dimensions, see for instance, Refs~\cite{Henning:2015alf,Henning:2017fpj,Gripaios:2018zrz, Criado:2019ugp, Criado:2019ugp, Fonseca:2019yya,Marinissen:2020jmb}. On the other hand, if there are new particles that have a mass less than $\LEW$ and are most likely a singlet under the SM gauge group, such as sterile neutrinos, they must be incorporated into the EFT framework thus extending the regime of SMEFT~\cite{Aparici:2009fh,
delAguila:2008ir,Bhattacharya:2015vja,Liao:2016qyd}.

Since many measurements are made below the electroweak scale, it is necessary to develop EFTs below $\LEW$. By integrating out the heavy particles in SM, i.e., the weak gauge bosons $W^\pm,~Z$, the Higgs boson $h$, and the top quark $t$, we arrive at the so-called low energy effective field theory (LEFT). It thus includes all other SM fields as its dynamical degrees of freedom including five quarks, all neutral and charged leptons, and respects the gauge symmetry $SU(3)_C\times U(1)_\EM$. It has been successfully applied in flavor physics; for a review, see for instance, Ref.~\cite{Buchalla:1995vs}. In recent years LEFT has been systematically developed. The classification of its basis operators up to dimension 6 and their tree-level and one-loop matching to the SMEFT also up to dimension 6 have been made in Refs.~\cite{Jenkins:2017jig,Dekens:2019ept}. (We note in passing that the basis of dim-6 operators in LEFT extended with light sterile neutrinos has been worked out recently~\cite{Chala:2020vqp,Li:2020lba}.) The complete one-loop renormalization of those basis operators has been accomplished in Ref.~\cite{Jenkins:2017dyc}. In this work we will push this systematic investigation one step further by building the basis of  dim-7 operators in LEFT and matching the effective interactions at tree level between SMEFT and LEFT both to dim-7 operators.

The outline of this paper is as follows. We first establish in section~\ref{sec2} the basis of dim-7 operators in LEFT, and then do the tree-level matching between the SMEFT and the LEFT in section~\ref{sec3} by incorporating new terms due to dim-7 operators in SMEFT or LEFT or in both. As a simple yet interesting application we study in section~\ref{sec4} the lepton number violating neutrino-photon interactions arising from dim-7 operators, and calculate various scattering cross sections and compare them with the SM results. We will also show a few examples of ultraviolet completion of a dim-7 operator in SMEFT that enters the above neutrino-photon interactions. Our main results are finally summarized in section~\ref{sec5}.

%%%%%%%%%%%%%%%%%%%%%%%
\section{The basis of dim-7 operator in LEFT}
\label{sec2}
%%%%%%%%%%%%%%%%%%%%%%%

In the LEFT where we are working the electroweak symmetry breakdown has already taken place, so that the gauge group is $SU(3)_C\times U(1)_\EM$. We have also integrated out the heavy particles of a mass of order $\LEW$, i.e., the weak gauge bosons $W^\pm,~Z$, the Higgs boson $h$, and the top quark $t$. Then the dynamical degrees of freedom are the $n_f=3$ number of the down-type quarks ($d,~s,~b$) and of the neutral ($\nu_{1,2,3}$) plus charged ($e,~\mu,~\tau$) leptons, the $n_u=2$ number of the up-type quarks ($u,~c$), and the photon ($A_\mu$) and eight gluons ($G_\mu^A$). Although we work with chiral fields ($\psi_{L,R}$), we assume they are already in their mass eigenstates. This means that any factors of quark and lepton mixing matrix elements are hidden in the Wilson coefficients of high dimensional operators. We label the fermion fields usually by the indices $p,~r,~s,~t$, i.e., $\nu_p,~e_{ip},~u_{ip},~d_{ip}$ with chirality $i=L,~R$, that appear in the same order in an operator and its Wilson coefficient. For specific applications these indices assume a generation value or a flavor name interchangeably.

The bases of dim-5 and dim-6 operators have been established in Ref.~\cite{Jenkins:2017jig}. In the following we will do the similar thing for dim-7 operators. First of all, Lorentz symmetry restricts dim-7 operators to the following possible classes:
\begin{align}
&\psi^2X^2,&
& \psi^4D,&
& \psi^2XD^2,&
& \psi^2D^4,
\end{align}
where the gauge covariant derivative is $D_\mu=\partial_\mu-ieQA_\mu-ig_sT^AG_\mu^A$ with $Q$ and $T^A$ being the electric charge and color generators and with $e$ and $g_s$ being gauge couplings, and $X_{\mu\nu}=F_{\mu\nu},~G_{\mu\nu}^A$ are the gauge field strength tensors. Note that there are no pure bosonic operators made out of $X$ and $D$ because Lorentz invariance requires an even number of $D$ factors which however cannot lead to an odd-dimensional operator. The operators in the last two classes $\psi^2XD^2$ and $\psi^2D^4$ are actually reducible to those in the first two classes $\psi^2X^2$ and $\psi^4D$ plus lower dimensional (i.e., dim-5 and dim-6) ones already covered in~\cite{Jenkins:2017jig}, by the use of equations of motion (EoMs) and integration by parts (IBP). Consider first the class $\psi^2D^4$. By Lorentz symmetry the two fermion fields must form a scalar $\bar\psi_1\psi_2$ or tensor $\bar\psi_1\sigma_{\mu\nu}\psi_2$ bilinear with all of the four factors of $D$ arranged by IBP to act on $\psi_2$. For the tensor bilinear, if $D^\mu$ and $D^\nu$ are adjacent, $\bar\psi_1\sigma_{\mu\nu}\cdots D^\mu D^\nu\cdots\psi_2$ reduces to the $\psi^2XD^2$ class by the relation $[D_\mu,D_\nu]\propto X_{\mu\nu}$, which will be coped with in a moment; otherwise, $D_\nu$ (or equivalently $D_\mu$) stays on the far right or far left. For the former, we proceed as $\overline{\psi_1}\cdots i\sigma_{\mu\nu}D^\nu\psi_2
=\overline{\psi_1}\cdots D_\mu\psi_2-\overline{\psi_1}\cdots \gamma_\mu\slashed{D}\psi_2$, where the second term yields by EoMs the lower dimensional operators covered in~\cite{Jenkins:2017jig} and the first belongs to the scalar bilinear that we will reduce further. If $D_\nu$ stays on the far left, we make it act on $\psi_1$ by IBP instead and then a similar manipulation to the above applies. The scalar bilinear is easy to handle. The four $D$ factors are contracted by either $\epsilon_{\mu\nu\rho\sigma}$ or $g_{\mu\nu}$, both of which reduce to the operators in the other three classes ($\psi^2X^2,~\psi^4D,~\psi^2XD^2$) and lower dimensional ones by the relations $[D_\mu,D_\nu]\propto X_{\mu\nu}$ and $D^2\psi=\slashed{D}\slashed{D}\psi={\rm EoM~ operators}$. This establishes the reducibility of the class $\psi^2D^4$.

Now we turn to reduce the class $\psi^2XD^2$. Again, the two fermion fields must form either a scalar or a tensor bilinear. The operators with a tensor bilinear can be transformed into those with a scalar bilinear plus EoM operators by the use of IBP, EoMs, and the Bianchi identity (BI) $D^\mu X^{\nu\rho}+D^\nu X^{\rho\mu}+D^\rho X^{\mu\nu}=0$. The proof goes like this. There are two types of Lorentz contractions: (a) $(\overline{\psi_1}\sigma_{\mu\nu}\psi_2)X^{\mu\nu}D^\rho D_\rho$ and (b) $(\overline{\psi_1}\sigma_{\mu\nu}\psi_2)X^{\nu\rho}D^\mu D_\rho$. By IBP we choose $X$ to be derivative free. Then we will not bother to display gauge group indices involved in $X$ which do not interrupt reduction of operators. For type (b), there are six ways of attaching $D^2$ to fermion fields. Two of them are reduced to scalar bilinear operators and EoM operators (shown as $\fbox{EoM}$):
\begin{eqnarray}
\nonumber
(\overline{D^\mu\psi_1}i\sigma_{\mu\nu}D_\rho \psi_2)X^{\nu\rho}&=&(\overline{D^\nu\psi_1}D_\rho\psi_2)X^{\nu\rho}
+(\overline{\psi_1}\overleftarrow{\slashed{D}}\gamma_\nu D_\rho\psi_2)X^{\nu\rho},
\\\nonumber
&\xlongrightarrow[]{\rm EoM}&(\overline{D^\nu\psi_1}D_\rho\psi_2)X^{\nu\rho}+\fbox{EoM},
\\\nonumber
(\overline{\psi_1}i\sigma_{\mu\nu}D^\mu D_\rho \psi_2)X^{\nu\rho}&=&(\overline{\psi_1}i\sigma_{\mu\nu}[D^\mu ,D_\rho] \psi_2)X^{\nu\rho}-{1\over 2}(\overline{\psi_1}[D_\rho, D_\nu]\psi_2)X^{\nu\rho}+(\overline{\psi_1}\gamma_\nu D_\rho\slashed{D}\psi_2)X^{\nu\rho}
\\
&\xlongrightarrow[]{\rm EoM}&\psi^2X^2 +\fbox{EoM},
\label{reduction}
\end{eqnarray}
and the other four with $D_\rho$ and $D^\mu$ interchanged or with both $D_\rho$ and $D^\mu$ acting on $\psi_1$ are similarly reduced. For type (a), excluding the trivial EoM operators with $D^2$ acting on a single fermion field, we are left with the following operator whose reduction goes as follows:
\begin{align}
\nonumber
(\overline{D_\rho\psi_1}i\sigma_{\mu\nu}D^\rho\psi_2)X^{\mu\nu}
&\xlongequal[]{\rm IBP}-(\overline{\psi_1}i\sigma_{\mu\nu}D_\rho\psi_2)D^\rho X^{\mu\nu}-(\overline{\psi_1}i\sigma_{\mu\nu}D^2\psi_2)X^{\mu\nu}
\\\nonumber
&\xlongequal[]{\rm BI}2(\overline{\psi_1}i\sigma_{\mu\nu}D_\rho\psi_2)D^\mu X^{\nu\rho}+\fbox{EoM}
\\
&\xlongequal[]{\rm IBP}-2(\overline{D^\mu\psi_1}i\sigma_{\mu\nu}D_\rho\psi_2)X^{\nu\rho}
-2(\overline{\psi_1}i\sigma_{\mu\nu} D^\mu D_\rho \psi_2)X^{\nu\rho}
+\fbox{EoM}.
\end{align}
The two operators on the right-hand side have already been reduced in equation~\eqref{reduction}. The reducibility of the class $\psi^2XD^2$ now rests on that of the scalar bilinear operators. In this case the Lorentz indices in $X$ and $D^2$ have to be contracted. Since $D_\mu X^{\mu\nu}$ only yields EoM operators that can be discarded as lower dimensional operators, we can apply IBP to make each fermion field be acted upon by one derivative. This yields the unique operator which itself is reducible:
\begin{eqnarray}
(\overline{D_\mu\psi_1}D_\nu \psi_2) X^{\mu\nu}
&\xlongequal[]{\rm IBP}&
-{1\over2}\overline{\psi_1}[D_\mu, D_\nu] \psi_2X^{\mu\nu}-\overline{\psi_1}(D_\nu \psi_2)D_\mu X^{\mu\nu}
+\fbox{T}
\nonumber
\\
&\xlongrightarrow[]{\rm EoM}&\psi^2X^2+\psi^4D,
\end{eqnarray}
where $\fbox{T}$ stands for the total derivative terms that can be discarded in the effective Lagrangian. This finally establishes the reducibility of the class $\psi^2XD^2$.

\begin{table}[!h]
\center
\begin{tabular}{|l | l |  l|l | l | l|}
\hline
 Operator & Specific form & $\#~(n_f,~n_u)$ &  Operator & Specific form & $\#~(n_f,~n_u)$
\\
\hline
\hline%%%%%
\multicolumn{6}{|c|}{{\color{magenta}$(\Delta L, \Delta B)=(0,~0)$}}
\\
\hline%%%
\multicolumn{3}{|c|}{\cellcolor{gray!25}$(\overline{R}L)X^2$}
&
\multicolumn{3}{|c|}{\cellcolor{gray!25}$(\overline{R}L)X\tilde{X}$}
\\
\hline%%
$\calO_{eF1}$ & $\alpha_{\rm em}(\overline{e_R}e_L) F_{\mu\nu}F^{\mu\nu}$ & $n_f^2$ &
$\calO_{eF2}$ & $\alpha_{\rm em}(\overline{e_R}e_L) F_{\mu\nu}\tilde{F}^{\mu\nu}$ & $n_f^2$
\\
$\calO_{dF1}$ & $\alpha_{\rm em}(\overline{d_R}d_L) F_{\mu\nu}F^{\mu\nu}$ & $n_f^2$ &
$\calO_{dF2}$ & $\alpha_{\rm em}(\overline{d_R}d_L) F_{\mu\nu}\tilde{F}^{\mu\nu}$ & $n_f^2$
\\
$\calO_{uF1}$ & $\alpha_{\rm em}(\overline{u_R}u_L) F_{\mu\nu}F^{\mu\nu}$ & $n_u^2$ &
$\calO_{uF2}$ & $\alpha_{\rm em}(\overline{u_R}u_L) F_{\mu\nu}\tilde{F}^{\mu\nu}$ & $n_u^2$
\\
$\calO_{dFG1}$ & $eg_3(\overline{d_R}T^A d_L) F_{\mu\nu}G^{A\mu\nu}$ & $n_f^2$ &
$\calO_{dFG2}$ & $eg_3(\overline{d_R}T^A d_L) F_{\mu\nu}\tilde{G}^{A\mu\nu}$ & $n_f^2$
\\
$\calO_{dFG3}$ & $eg_3(\overline{d_R}T^A\sigma^{\mu\nu} d_L) F_{\mu\rho}G^{A\rho}_\nu$ & $n_f^2$ & & &
\\
$\calO_{uFG1}$ & $eg_3(\overline{u_R}T^A u_L) F_{\mu\nu}G^{A\mu\nu}$ & $n_u^2$ &
$\calO_{uFG2}$ & $eg_3(\overline{u_R}T^A u_L) F_{\mu\nu}\tilde{G}^{A\mu\nu}$ & $n_u^2$
\\
$\calO_{uFG3}$ & $eg_3(\overline{u_R}T^A\sigma^{\mu\nu} u_L) F_{\mu\rho}G^{A\rho}_\nu$ & $n_u^2$ & & &
\\
$\calO_{eG1}$ & $\alpha_s(\overline{e_R}e_L) G^A_{\mu\nu}G^{A\mu\nu}$ & $n_f^2$ &
$\calO_{eG2}$ & $\alpha_s(\overline{e_R}e_L) G^A_{\mu\nu}\tilde{G}^{A\mu\nu}$ & $n_f^2$
\\
$\calO_{dG1}$ & $\alpha_s(\overline{d_R}d_L) G^A_{\mu\nu}G^{A\mu\nu}$ & $n_f^2$ &
$\calO_{dG2}$ & $\alpha_s(\overline{d_R}d_L) G^A_{\mu\nu}\tilde{G}^{A\mu\nu}$ & $n_f^2$
\\
$\calO_{dG3}$ & $\alpha_sd_{ABC}(\overline{d_R}T^Ad_L) G^B_{\mu\nu}G^{C\mu\nu}$ & $n_f^2$ &
$\calO_{dG4}$ & $\alpha_sd_{ABC}(\overline{d_R}T^Ad_L) G^B_{\mu\nu}\tilde{G}^{C\mu\nu}$ & $n_f^2$
\\
$\calO_{dG5}$ & $\alpha_sf_{ABC}(\overline{d_R}T^A\sigma^{\mu\nu}d_L) G^B_{\mu\rho}G^{C\rho}_\nu$ & $n_f^2$ & & &
\\
$\calO_{uG1}$ & $\alpha_s(\overline{u_R}u_L) G^A_{\mu\nu}G^{A\mu\nu}$ & $n_u^2$ &
$\calO_{uG2}$ & $\alpha_s(\overline{u_R}u_L) G^A_{\mu\nu}\tilde{G}^{A\mu\nu}$ & $n_u^2$
\\
$\calO_{uG3}$ & $\alpha_sd_{ABC}(\overline{u_R}T^Au_L) G^B_{\mu\nu}G^{C\mu\nu}$ &  $n_u^2$ &
$\calO_{uG4}$ & $\alpha_sd_{ABC}(\overline{u_R}T^Au_L) G^B_{\mu\nu}\tilde{G}^{C\mu\nu}$ & $n_u^2$
\\
$\calO_{uG5}$ & $\alpha_sf_{ABC}(\overline{u_R}T^A\sigma^{\mu\nu}u_L) G^B_{\mu\rho}G^{C\rho}_\nu$ & $n_u^2$ & & &
\\
\hline%%
\multicolumn{3}{|c|}{\cellcolor{gray!25}$(\bar{L}\gamma^\mu L) (\bar{L}\overleftrightarrow{D_\mu} R)$}
&
\multicolumn{3}{|c|}{\cellcolor{gray!25}$(\bar{R}\gamma^\mu R) (\bar{L}\overleftrightarrow{D_\mu} R)$}
\\
\hline%%
$\calO_{\nu eD}$ & $(\overline{\nu}\gamma^\mu \nu)(\overline{e_L} i\overleftrightarrow{D_\mu} e_R)$ & $n_f^4$ & & &
\\
$\calO_{\nu dD}$ & $(\overline{\nu}\gamma^\mu \nu)(\overline{d_L} i\overleftrightarrow{D_\mu} d_R)$ & $n_f^4$ & & &
\\
$\calO_{\nu uD}$ & $(\overline{\nu}\gamma^\mu \nu)(\overline{u_L} i\overleftrightarrow{D_\mu} u_R)$ & $n_f^2n_u^2$ & & &
\\
$\calO_{eeD1}$ & $(\overline{e_L}\gamma^\mu e_L)(\overline{e_L} i\overleftrightarrow{D_\mu} e_R)$ & $\frac{1}{2}n_f^3(n_f-1)$ &
$\calO_{eeD2}$ & $(\overline{e_R}\gamma^\mu e_R)(\overline{e_L} i\overleftrightarrow{D_\mu} e_R)$ & $\frac{1}{2}n_f^3(n_f-1)$
\\
$\calO_{edD1}$ & $(\overline{e_L}\gamma^\mu e_L)(\overline{d_L} i\overleftrightarrow{D_\mu} d_R)$ & $n_f^4$ &
$\calO_{edD2}$ & $(\overline{e_R}\gamma^\mu e_R)(\overline{d_L} i\overleftrightarrow{D_\mu} d_R)$ & $n_f^4$
\\
$\calO_{euD1}$ & $(\overline{e_L}\gamma^\mu e_L)(\overline{u_L} i\overleftrightarrow{D_\mu} u_R)$ & $n_f^2n_u^2$ &
$\calO_{euD2}$ & $(\overline{e_R}\gamma^\mu e_R)(\overline{u_L} i\overleftrightarrow{D_\mu} u_R)$ & $n_f^2n_u^2$
\\
$\calO_{deD1}$ & $(\overline{d_L}\gamma^\mu d_L)(\overline{e_L} i\overleftrightarrow{D_\mu} e_R)$ & $n_f^4$ &
$\calO_{deD2}$ & $(\overline{d_R}\gamma^\mu d_R)(\overline{e_L} i\overleftrightarrow{D_\mu} e_R)$ & $n_f^4$
\\
$\calO_{ddD1}$ & $(\overline{d_L}\gamma^\mu d_L)(\overline{d_L} iD_\mu d_R)$ & $n_f^4$ &
$\calO_{ddD2}$ & $(\overline{d_R}\gamma^\mu d_R)(\overline{d_L} i\overleftrightarrow{D_\mu} d_R)$ & $n_f^4$
\\
$\calO_{duD1}$ & $(\overline{d_L}\gamma^\mu d_L)(\overline{u_L} i\overleftrightarrow{D_\mu} u_R)$ & $n_f^2n_u^2$ &
$\calO_{duD3}$ & $(\overline{d_R}\gamma^\mu d_R)(\overline{u_L} \overleftrightarrow{D_\mu} u_R)$ & $n_f^2n_u^2$
\\
$\calO_{duD2}$ & $(\overline{d_L}\gamma^\mu d_L][\overline{u_L} i\overleftrightarrow{D_\mu} u_R)$ & $n_f^2n_u^2$ &
$\calO_{duD4}$ & $(\overline{d_R}\gamma^\mu d_R][\overline{u_L} i\overleftrightarrow{D_\mu} u_R)$ & $n_f^2n_u^2$
\\
$\calO_{ueD1}$ & $(\overline{u_L}\gamma^\mu u_L)(\overline{e_L} i\overleftrightarrow{D_\mu} e_R)$ & $n_f^2n_u^2$ &
$\calO_{ueD2}$ & $(\overline{u_R}\gamma^\mu u_R)(\overline{e_L} i\overleftrightarrow{D_\mu} e_R)$ & $n_f^2n_u^2$
\\
$\calO_{udD1}$ & $(\overline{u_L}\gamma^\mu u_L)(\overline{d_L} i\overleftrightarrow{D_\mu} d_R)$ & $n_f^2n_u^2$ &
$\calO_{udD3}$ & $(\overline{u_R}\gamma^\mu u_R)(\overline{d_L} i\overleftrightarrow{D_\mu} d_R)$ & $n_f^2n_u^2$
\\
$\calO_{udD2}$ & $(\overline{u_L}\gamma^\mu u_L] [\overline{d_L} i\overleftrightarrow{D_\mu} d_R)$ & $n_f^2n_u^2$ &
$\calO_{udD4}$ & $(\overline{u_R}\gamma^\mu u_R][\overline{d_L} i\overleftrightarrow{D_\mu} d_R)$ & $n_f^2n_u^2$
\\
$\calO_{uuD1}$ & $(\overline{u_L}\gamma^\mu u_L)(\overline{u_L} i\overleftrightarrow{D_\mu} u_R)$ & $n_u^4$ &
$\calO_{uuD2}$ & $(\overline{u_R}\gamma^\mu u_R)(\overline{u_L} i\overleftrightarrow{D_\mu} u_R)$ & $n_u^4$
\\
$\calO_{\nu eduD}$ & $(\overline{\nu}\gamma^\mu e_L)(\overline{d_L} i\overleftrightarrow{D_\mu} u_R)$ & $n_f^3n_u$ & & &
\\
$\calO_{e\nu udD}$ & $(\overline{e_L}\gamma^\mu \nu)(\overline{u_L} i\overleftrightarrow{D_\mu} d_R)$ & $n_f^3n_u$ & & &
\\
$\calO_{du\nu eD1}$ & $(\overline{d_L}\gamma^\mu u_L)(\overline{\nu}   i\overleftrightarrow{D_\mu} e_R)$ & $n_f^3n_u$ &
$\calO_{du\nu eD2}$ & $(\overline{d_R}\gamma^\mu u_R)(\overline{\nu}i\overleftrightarrow{D_\mu} e_R)$ & $n_f^3n_u$
\\
\hline%%
\multicolumn{6}{|c|}{$\#$ total: $9n_f^4+n_f^3(4n_u-1)+n_f^2(13n_u^2+14)+2n_u^2(n_u^2+5)~
(\Rightarrow 1584\textrm{ at }n_u=2,~n_f=3)$}
\\
\hline
\end{tabular}
\caption{Basis of dim-7 operators in LEFT with lepton and baryon numbers conserved, i.e., $\Delta B=\Delta L=0$, together with count of operators for general $n_u,~n_f$. $L,~R$ refer to left- and right-handed fermion fields, and $\alpha_{\rm em}=e^2/(4\pi)$ and $\alpha_s=g_s^2/(4\pi)$. The two brackets $(,)$ and $[,]$ indicate two different color contractions. Hermitian conjugate operators are not displayed.}
\label{tab1}
\end{table}

%%%%table
\begin{table}[!h]
\center
\begin{tabular}{|l | l |  l|l | l | l|}
\hline
 Operator & Specific form & $\#~(n_f,~n_u)$ &  Operator & Specific form & $\#~(n_f,~n_u)$
\\
\hline
\hline%%%%%
\multicolumn{6}{|c|}{{\color{magenta}$(\Delta L, \Delta B)=(2,~0)$}}
\\
\hline%%%
\multicolumn{3}{|c|}{\cellcolor{gray!25}$(\overline{L^C}L)X^2$}
&
\multicolumn{3}{|c|}{\cellcolor{gray!25}$\overline{L^C}L)X\tilde{X}$}
\\
\hline%%
$\calO_{\nu F1}$ & $\alpha_{\rm em}(\overline{\nu^C}\nu)F_{\mu\nu}F^{\mu\nu}$ & $\frac{1}{2}n_f(n_f+1)$ &
$\calO_{\nu F2}$ & $\alpha_{\rm em}(\overline{\nu^C}\nu)F_{\mu\nu}\tilde{F}^{\mu\nu}$ & $\frac{1}{2}n_f(n_f+1)$
\\
$\calO_{\nu G1}$ & $\alpha_s(\overline{\nu^C}\nu)G^A_{\mu\nu}G^{A\mu\nu}$ & $\frac{1}{2}n_f(n_f+1)$ &
$\calO_{\nu G2}$ & $\alpha_s(\overline{\nu^C}\nu)G^A_{\mu\nu}\tilde{G}^{A\mu\nu}$ & $\frac{1}{2}n_f(n_f+1)$
\\
\hline%%%
\multicolumn{3}{|c|}{\cellcolor{gray!25}$(\bar{L}\gamma^\mu L)(\overline{L^C}\overleftrightarrow{D_\mu}L)$}
&
\multicolumn{3}{|c|}{\cellcolor{gray!25}$(\bar{R}\gamma^\mu R)(\overline{L^C}\overleftrightarrow{D_\mu}L)$}
\\
\hline%%
$\calO_{\nu\nu D}$ & $(\overline{\nu}\gamma^\mu \nu)(\overline{\nu^C}i\overleftrightarrow{\partial_\mu} \nu)$ & $\frac{1}{6}n_f^2(n_f^2-3n_f+2)$ & & &
\\
$\calO_{e\nu D1}$ & $(\overline{e_L}\gamma^\mu e_L)(\overline{\nu^C} i\overleftrightarrow{\partial_\mu} \nu)$ & $\frac{1}{2}n_f^3(n_f-1)$ &
$\calO_{e\nu D2}$ & $(\overline{e_R}\gamma^\mu e_R)(\overline{\nu^C} i\overleftrightarrow{\partial_\mu} \nu)$ & $\frac{1}{2}n_f^3(n_f-1)$
\\%%
$\calO_{d\nu D1}$ & $(\overline{d_L}\gamma^\mu d_L)(\overline{\nu^C} i\overleftrightarrow{\partial_\mu} \nu)$ & $\frac{1}{2}n_f^3(n_f-1)$ &
$\calO_{d\nu D2}$ & $(\overline{d_R}\gamma^\mu d_R)(\overline{\nu^C} i\overleftrightarrow{\partial_\mu} \nu)$ & $\frac{1}{2}n_f^3(n_f-1)$
\\
$\calO_{u\nu D1}$ & $(\overline{u_L}\gamma^\mu u_L)(\overline{\nu^C} i\overleftrightarrow{\partial_\mu} \nu)$ & $\frac{1}{2}n_fn_u^2(n_f-1)$ &
$\calO_{u\nu D2}$ & $(\overline{u_R}\gamma^\mu u_R)(\overline{\nu^C} i\overleftrightarrow{\partial_\mu} \nu)$ & $\frac{1}{2}n_fn_u^2(n_f-1)$
\\
$\calO_{due\nu D1}$ & $(\overline{d_L}\gamma^\mu u_L)(\overline{e_L^C} i\overleftrightarrow{D_\mu} \nu)$ & $n_f^3n_u$ &
$\calO_{ due\nu D2}$ & $(\overline{d_R}\gamma^\mu u_R)(\overline{e_L^C} i\overleftrightarrow{D_\mu} \nu)$ & $n_f^3n_u$
\\
\hline%%%
\multicolumn{3}{|c|}{\cellcolor{gray!25}$(\overline{R^C}\gamma^\mu L)(\bar{R}\overleftrightarrow{D_\mu}L)$}
&
\multicolumn{3}{|c|}{\cellcolor{gray!25}$(\overline{R^C}\gamma^\mu L)(\bar{L}\overleftrightarrow{D_\mu}R)$}
\\
\hline%%
$\calO_{ e\nu udD1}$ & $(\overline{e_R^C} \gamma^\mu \nu )(\overline{d_R} i\overleftrightarrow{D_\mu} u_L)$ & $n_f^3n_u$  &
$\calO_{e\nu udD2}$ & $(\overline{e_R^C} \gamma^\mu \nu )(\overline{d_L} i\overleftrightarrow{D_\mu} u_R)$ & $n_f^3n_u$
\\
\hline
\multicolumn{6}{|c|}{$\#$ total: $\frac{1}{6}n_f\big( 13n_f^3+3n_f^2(8n_u-5)+2n_f(3n_u^2+7)-6n_u^2+12\big)~
(\Rightarrow 375\textrm{ at }n_u=2,~n_f=3)$}
\\
\hline%%%%%
 \multicolumn{6}{|c|}{{\color{magenta}$(\Delta L, \Delta B)=(1,~-1)$}}
\\
\hline%%%
\multicolumn{3}{|c|}{\cellcolor{gray!25}$(\bar{L}\gamma^\mu L)(\bar{L}\overleftrightarrow{D_\mu}L^C)$}
&
\multicolumn{3}{|c|}{\cellcolor{gray!25}$(\bar{L}\gamma^\mu L)(\bar{R}\overleftrightarrow{D_\mu}R^C)$}
\\
\hline%%
& & &
$\calO_{d\nu udD1}$ & $(\overline{d_L}\gamma^\mu \nu)(\overline{u_R} i\overleftrightarrow{D_\mu} d_R^{C})$ & $n_f^3n_u$
\\
$\calO_{u\nu dD1}$ & $(\overline{u_L}\gamma^\mu \nu)(\overline{d_L} i\overleftrightarrow{D_\mu} d_L^{C})$ & $\frac{1}{2}n_f^2n_u(n_f+1)$ &
$\calO_{u\nu dD2}$ & $(\overline{u_L}\gamma^\mu \nu)(\overline{d_R} i\overleftrightarrow{D_\mu} d_R^{C})$ & $\frac{1}{2}n_f^2n_u(n_f+1)$
\\
$\calO_{dedD1}$ & $(\overline{d_L}\gamma^\mu e_L)(\overline{d_L} i\overleftrightarrow{D_\mu} d_L^{C})$ & $\frac{1}{6}n_f^2(n_f^2+3n_f+2)$ &
$\calO_{dedD2}$ & $(\overline{d_L}\gamma^\mu e_L)(\overline{d_R} i\overleftrightarrow{D_\mu} d_R^{C})$ & $\frac{1}{2}n_f^3(n_f+1)$
\\
\hline%%%
\multicolumn{3}{|c|}{\cellcolor{gray!25}$(\bar{R}\gamma^\mu R)(\bar{L}\overleftrightarrow{D_\mu}L^C)$}
&
\multicolumn{3}{|c|}{\cellcolor{gray!25}$(\bar{R}\gamma^\mu R)(\bar{R}\overleftrightarrow{D_\mu}R^C)$}
\\
\hline%%
$\calO_{dedD3}$ & $(\overline{d_R}\gamma^\mu e_R)(\overline{d_L} i\overleftrightarrow{D_\mu} d_L^{C})$ & $\frac{1}{2}n_f^3(n_f+1)$ &
$\calO_{dedD4}$ & $(\overline{d_R}\gamma^\mu e_R)(\overline{d_R} i\overleftrightarrow{D_\mu} d_R^{C})$& $\frac{1}{6}n_f^2(n_f^2+3n_f+2)$
\\
\hline
 \multicolumn{6}{|c|}{$\#$ total: $\frac{1}{3}n_f^2(2n_f+1)(2n_f+3n_u+2)~
 (\Rightarrow 294\textrm{ at }n_u=2,~n_f=3)$}
\\
\hline%%%%%
\multicolumn{6}{|c|}{{\color{magenta}$(\Delta L, \Delta B)=(1,~1)$}}
\\
\hline%%%
\multicolumn{3}{|c|}{\cellcolor{gray!25}$(\overline{R^C}\gamma^\mu L)(\overline{L^C}\overleftrightarrow{D_\mu}L)$}
&
\multicolumn{3}{|c|}{\cellcolor{gray!25}$(\overline{R^C}\gamma^\mu L)(\overline{R^C}\overleftrightarrow{D_\mu}R)$}
\\
\hline%%
$\calO_{d\nu udD2}$ & $(\overline{d_R^C} \gamma^\mu \nu)(\overline{u_L^C} i\overleftrightarrow{D_\mu} d_L)$ & $n_f^3n_u$ & & &
\\
$\calO_{u\nu dD3}$ & $(\overline{u_R^C} \gamma^\mu \nu)(\overline{d_L^C} i\overleftrightarrow{D_\mu} d_L)$ & $\frac{1}{2}n_f^2n_u(n_f+1)$ &
$\calO_{u\nu dD4}$ & $(\overline{u_R^C} \gamma^\mu \nu)(\overline{d_R^C} i\overleftrightarrow{D_\mu}  d_R)$ & $\frac{1}{2}n_f^2n_u(n_f+1)$
\\
$\calO_{deuD1}$ & $(\overline{d_R^C} \gamma^\mu e_L)(\overline{u_L^C} i\overleftrightarrow{D_\mu}  u_L)$ & $\frac{1}{2}n_f^2n_u(n_u+1)$ &
$\calO_{deuD2}$ & $(\overline{d_R^C} \gamma^\mu e_L)(\overline{u_R^C} i\overleftrightarrow{D_\mu}  u_R)$ & $\frac{1}{2}n_f^2n_u(n_u+1)$
\\
\cline{4-6}
$\calO_{ueudD1}$ & $(\overline{u_R^C} \gamma^\mu e_L)(\overline{u_L^C} i\overleftrightarrow{D_\mu}  d_L)$ & $n_f^2n_u^2$ &
\multicolumn{3}{|c|}{\cellcolor{gray!25}$(\overline{L^C}\gamma^\mu R)(\overline{R^C}\overleftrightarrow{D_\mu}R)$}
\\
\hline%%%
\multicolumn{3}{|c|}{\cellcolor{gray!25}$(\overline{L^C}\gamma^\mu R)(\overline{L^C}\overleftrightarrow{D_\mu}L)$}
&
$\calO_{deuD4}$ & $(\overline{d_L^C} \gamma^\mu e_R)(\overline{u_R^C} i\overleftrightarrow{D_\mu}  u_R)$ & $\frac{1}{2}n_f^2n_u(n_u+1)$
\\
\cline{1-3}%%%
$\calO_{deuD3}$ & $(\overline{d_L^C} \gamma^\mu e_R)(\overline{u_L^C} i\overleftrightarrow{D_\mu}  u_L)$ & $\frac{1}{2}n_f^2n_u(n_u+1)$ &
$\calO_{ueduD2}$ & $(\overline{u_L^C}\gamma^\mu e_R)(\overline{u_R^C} i\overleftrightarrow{D_\mu}  d_R)$ & $n_f^2n_u^2$
 \\
\hline
\multicolumn{6}{|c|}{$\#$ total: $9n_f^4+n_f^3(4n_u-1)+n_f^2(13n_u^2+14)+2n_u^2(n_u^2+5)~
(\Rightarrow 306\textrm{ at }n_u=2,~n_f=3)$}
\\
\hline
\end{tabular}
\caption{Basis of dim-7 operators in LEFT with lepton or baryon number or both violated, i.e., $(\Delta B,\Delta L)=(2,0),~(1,-1),~(1,1)$, together with count of operators for general $n_u,~n_f$. Color contraction is implied for triple quark fields, and $\tilde H=i\sigma_2H^*$ and $iD_\mu\psi^C\equiv (iD_\mu \psi)^C$ for brevity. Hermitian conjugate operators are not displayed.}
\label{tab2}
\end{table}

We are thus left with the classes $\psi^2X^2$ and $\psi^4D$ to examine further. Working with chiral fermion fields we see that the monomial operators in these classes are all non-Hermitian due to the special Lorentz structure which can be formed. So in the following we will work out one half of them while the other half can be obtained by Hermitian conjugate. We start with the class $\psi^2X^2$, which may take the following forms for generic chiral fermion fields $\psi_{1,2}$:
\begin{align}
&\calO_{\psi F1}=(\overline{\psi_1}\psi_2)F_{\mu\nu}F^{\mu\nu}, &
&\calO_{\psi F2}=(\overline{\psi_1}\psi_2)F_{\mu\nu}\tilde{F}^{\mu\nu},
\nonumber
%\label{psiF}
\\
&\calO_{\psi FG1}=(\overline{\psi_1}T^A\psi_2)F_{\mu\nu}G^{A\mu\nu}, &
&\calO_{\psi FG2}=(\overline{\psi_1}T^A\psi_2)F_{\mu\nu}\tilde{G}^{A\mu\nu},
\nonumber
\\
&\calO_{\psi FG3}=
(\overline{\psi_1}T^A\sigma_{\mu\nu}\psi_2)F^{\mu\alpha}G^{A\nu}_\alpha,
\nonumber
\\
&\calO_{\psi G1}=(\overline{\psi_1}\psi_2)G^A_{\mu\nu}G^{A\mu\nu},&
&\calO_{\psi G2}=
(\overline{\psi_1}\psi_2)G^A_{\mu\nu}\tilde{G}^{A\mu\nu},
\nonumber
\\
&\calO_{\psi G3}=d_{ABC}(\overline{\psi_1}T^A\psi_2)G^B_{\mu\nu}G^{C\mu\nu}, &
&\calO_{\psi G4}=d_{ABC}(\overline{\psi_1}T^A\psi_2)G^B_{\mu\nu}\tilde{G}^{C\mu\nu},~~
\nonumber
\\
&\calO_{\psi G5}=
f_{ABC}(\overline{\psi_1}\sigma_{\mu\nu}T^A\psi_2)
G^{B\mu\alpha}G^{C\nu}_\alpha,
\label{psiG}
\end{align}
where the field strength dual $\tilde X^{\mu\nu}=\epsilon^{\mu\nu\rho\sigma}X_{\rho\sigma}/2$, $f_{ABC}$ is the structure constant of $SU(3)$, and $d_{ABC}$ the symmetric invariant appearing in the anticommutator of generators in the fundamental representation, $\{T^A,T^B\}=\delta^{AB}/3+d^{ABC}T^C$. The other possible operators either vanish or can be reduced to the above ones,
\begin{align}
&(\overline{\psi_1}\sigma_{\mu\nu}\psi_2)X^{\mu\alpha}X^\nu_{~\alpha}=0, &
&(\overline{\psi_1}\sigma_{\mu\nu}\psi_2)X^{\mu\alpha}
\tilde{X}^\nu_{~\alpha}
=0,~~~X=F,G^A,
\nonumber
\\
&(\overline{\psi_1}T^A\sigma_{\mu\nu}P_{\pm}\psi_2)
F^{\mu\alpha}\tilde{G}^{A\nu}_{~~~\alpha}=\pm i \calO_{\psi FG3}, &
&f_{ABC}(\overline{\psi_1}\sigma_{\mu\nu}T^AP_{\pm}\psi_2)
G^{B\mu\alpha}\tilde{G}^{C\nu}_{~~~\alpha}=\pm i\calO_{\psi G5},
\nonumber
\\
&f_{ABC}(\overline{\psi_1}T^A\psi_2)G^B_{\mu\nu}G^{C\mu\nu}=0, &
&f_{ABC}(\overline{\psi_1}T^A\psi_2)G^B_{\mu\nu}\tilde{G}^{C\mu\nu}=0,
\label{red:psiFG}
\end{align}
where the chiral projectors $P_\pm=(1\pm \gamma^5)/2$ are also understood to appear in $\calO_{\psi FG3}$ and $\calO_{\psi G5}$ of equation~\eqref{red:psiFG}. The above reduction makes use of the following identities,
\begin{align}
&\sigma_{\mu\nu}P_\pm=
\mp\frac{i}{2}\epsilon_{\mu\nu\rho\sigma}\sigma^{\rho\sigma}P_\pm,&
&\epsilon_{\mu\nu\rho\sigma}\epsilon^{\alpha\beta\gamma\sigma}=
-g^{[\alpha}_\mu g^\beta_\nu g^{\gamma]}_\rho,
\end{align}
with $[\dots]$ indicating antisymmetrization of the arguments inside. With equation~\eqref{psiG} it is easy to figure out the relevant fields $\psi_{1,2}$ and find the complete set of operators in this class. These operators conserve baryon number ($\Delta B=0$) but can conserve ($\Delta L=0$) or violate lepton number by two units ($\Delta L=\pm 2$), which are displayed respectively in tables~\ref{tab1} and \ref{tab2}.

For the class $\psi^4D$, there are two possible Lorentz structures,
\begin{align}
&(\overline{\psi_1}\sigma^{\mu\nu}\psi_2)
(\overline{\psi_3}\gamma_{[\mu}\overleftrightarrow{D_{\nu]}}\psi_4),&
&(\overline{\psi_1}\gamma^\mu\psi_2)
(\overline{\psi_3} i\overleftrightarrow{D_\mu}\psi_4),
\label{4phiD}
\end{align}
where $\overline{A}\overleftrightarrow{D_\mu}B=\overline{A}D_\mu B-\overline{A}\overleftarrow{D_\mu}B$. However, the two structures are not independent as the tensor structure can be reduced to the vector one plus dim-6 operators ($\fbox{dim-6}$) with the aid of EoMs, IBP, and the Fierz identities (FI):
\begin{eqnarray}
&&(\overline{\psi_1}\sigma^{\mu\nu}\psi_2)
(\overline{\psi_3}\gamma_{[\mu}i\overleftrightarrow{D_{\nu]}}\psi_4)
\nonumber
\\
&\xlongequal{\rm IBP}&
+2iD_\nu(\overline{\psi_1}\sigma^{\mu\nu}\psi_2)
(\overline{\psi_3}\gamma_\mu\psi_4)
+4(\overline{\psi_1}\sigma^{\mu\nu}\psi_2)
(\overline{\psi_3}\gamma_\mu iD_\nu\psi_4)+\fbox{\rm T}
\nonumber
\\
&\xlongequal{\rm EoM}&
-2(\overline{\psi_1}i\overleftrightarrow{D_\mu}\psi_2)
(\overline{\psi_3}\gamma^\mu\psi_4)
+4(\overline{\psi_1}\gamma^\mu\gamma^\nu \psi_2)
(\overline{\psi_3}\gamma_\mu iD_\nu \psi_4)+\fbox{dim-6}
\nonumber
\\
&\xlongequal[\rm EoM]{\rm FI,~IBP}&
\begin{cases}
-2(\overline{\psi_1}i\overleftrightarrow{D_\mu}P_\pm\psi_2)
(\overline{\psi_3}\gamma^\mu P_\pm\psi_4)
-4(\overline{\psi_1}i\overleftrightarrow{D_\mu}P_\pm \psi_4)
(\overline{\psi_3} \gamma^\mu P_\pm\psi_2)+\fbox{T}+\fbox{dim-6},
\\
-2(\overline{\psi_1}i\overleftrightarrow{D_\mu}P_\pm\psi_2)
(\overline{\psi_3}\gamma^\mu P_\mp\psi_4)+\fbox{dim-6}.
\end{cases}
\end{eqnarray}
In the second step we have used the relation $\sigma^{\mu\nu}=i\gamma^\mu\gamma^\nu-ig^{\mu\nu}
=ig^{\mu\nu}-i\gamma^\nu\gamma^\mu$, and in the last step distinguished between the two cases in which $\psi_{2,4}$ have the same or opposite chirality to apply the FIs:
\begin{align}\nonumber
(\overline{\psi_1}\gamma^\mu \gamma^\nu P_\pm\psi_2)(\overline{\psi_3}\gamma_\mu iD_\nu P_\pm\psi_4)=&-2(\overline{\psi_1}iD_\mu P_\pm\psi_4)(\overline{\psi_3}\gamma^\mu P_\pm\psi_2),
\\
(\overline{\psi_1}\gamma^\mu \gamma^\nu P_\pm\psi_2)(\overline{\psi_3}\gamma_\mu iD_\nu P_\mp\psi_4)=&
2(\overline{\psi_1}P_\pm\psi_2)(\overline{\psi_3}i\slashed{D} P_\mp\psi_4)
+2(\overline{\psi_1}i\slashed{D} P_\mp\psi_4)(\overline{\psi_3}P_\pm\psi_2).
\end{align}
Therefore, the tensor structure can be discarded in favor of the vector one in equation~\eqref{4phiD} to work out all possible operators. For a given field configuration $(\psi_1,~\psi_2,~\psi_3,~\psi_4)$ fulfilling gauge invariance, we may form several apparently different operators. However, we find there is only one independent operator by the use of IBP, EoM, and the following Fierz transformations~\cite{Liao:2016hru}:
\begin{align}\nonumber
-(\overline{\psi_1}\gamma^\mu P_\pm \psi_2)(\overline{\psi_3}P_\pm\psi_4)
=&(\overline{\psi_1}\gamma^\mu P_\pm\psi_3^C)(\overline{\psi_2^C}P_\pm\psi_4)
+(\overline{\psi_1}\gamma^\mu P_\pm \psi_4)(\overline{\psi_3}P_\pm\psi_2),
\\
-(\overline{\psi_1}\gamma^\mu P_\pm \psi_2)(\overline{\psi_3}P_\mp \psi_4)
=&(\overline{\psi_1}P_\mp \psi_3^C)(\overline{\psi_4^C}\gamma^\mu P_\pm\psi_2)
+(\overline{\psi_1} P_\mp\psi_4)(\overline{\psi_3}\gamma^\mu P_\pm\psi_2),
\end{align}
where charge conjugation is defined as $\psi^C=C\bar\psi^{\T}$ with the matrix $C$ satisfying the relations $C^{\T}=C^\dagger=-C$ and $C^2=-1$ so that $(\psi^C)^C=\psi$. Considering the above reduction, for a given configuration of fields $\psi_{1,2,3,4}\in \{u_{L/R},d_{L/R},e_{L/R},\nu \}$, one can write down the corresponding gauge invariant operators. The final complete operators in this class are given in the rest part of tables~\ref{tab1} and \ref{tab2} according to their lepton and baryon numbers.

In tables~\ref{tab1} and \ref{tab2} we also count the number of each operator for generally $n_u$ up-type quarks, $n_f$ down-type quarks, and $n_f$ neutral and charged leptons. Comparing to dim-7 operators in SMEFT~\cite{Lehman:2014jma,Liao:2016hru} and its sterile neutrino extended $\nu$SMEFT~\cite{Liao:2016qyd} which only have $(\Delta L, \Delta B)=(2,0),~(1,-1)$, the dim-7 operators in LEFT have additional sectors with $(\Delta L,\Delta B)=(2,0),~(1,1)$. In counting independent operators in each sector we have taken into account symmetries in their flavor indices. In the sector with $(\Delta L,\Delta B)=(0,0)$, only the operators $\calO_{eeD1}^{prst}$ and $\calO_{eeD2}^{prst}$ have flavor symmetries, and are respectively antisymmetric under $p\leftrightarrow s$ and $r\leftrightarrow t$ up to dim-6 terms by EoM, thus reducing the number of their independent operators. In the sector with $(\Delta L,\Delta B)=(2,0)$, the operators $\calO_{\nu F1,2}^{pr}$ and $\calO_{\nu G1,2}^{pr}$ are symmetric under $p\leftrightarrow r$, and $\calO_{e\nu D1,2}^{prst}$, $\calO_{d\nu D1,2}^{prst}$ and $\calO_{u\nu D1,2}^{prst}$ are antisymmetric in the neutrino indices $s,~t$, while $\calO_{\nu\nu D}^{prst}$ are totally antisymmetric in the neutrino indices $r,~s,~t$. In the sector with $(\Delta L,\Delta B)=(1,-1)$, the operators ${\cal O}_{u\nu dD1,2}^{prst}$ and ${\cal O}_{dedD2,3}^{prst}$ are symmetric under $s\leftrightarrow t$, while ${\cal O}_{dedD1,4}^{prst}$ are totally symmetric in $p,s,t$ for the three down-type quark fields up to dim-6 terms by EoM. In the last sector with $(\Delta L, \Delta B)=(1,1)$, the operators ${\cal O}_{u\nu dD3,4}^{prst}$ and ${\cal O}_{deuD1,2,3,4}^{prst}$ are all symmetric under $s\leftrightarrow t$. We have confirmed our above count of independent operators by the Hilbert series method in Ref.~\cite{Henning:2015alf}. By utilizing the {\em Mathematica} code developed in that reference, we generate all possible field configurations that can form a gauge and Lorentz invariant dim-7 operator, and count the total number of independent operators for each field configuration. This counting is easily done with $n_u$ up-type quarks, $n_f$ down-type quarks, and $n_f$ neutral and charged leptons, and is in accord with the counts shown in tables~\ref{tab1} and \ref{tab2} obtained by an analysis of flavor symmetries. For the SM case with $n_u=2,~n_f=3$ and including Hermitian conjugates of the operators, there are in total $3168|^{\Delta L=0}_{\Delta B=0}+750|^{\Delta L=\pm 2}_{\Delta B=0}+588|^{L=\pm1}_{\Delta B=\mp1}+712|^{\Delta L=\pm1}_{\Delta B=\pm1}$ independent dim-7 operators in LEFT.

%%%%%%%%%%%%%%%%%%%%%%%
\section{Matching with SMEFT up to dimension 7}
%%%%%%%%%%%%%%%%%%%%%%%
\label{sec3}

Although the SMEFT is defined above the electroweak scale $\LEW$ and stays closer to certain new physics at the scale $\LNP$, we have to employ the LEFT defined below $\LEW$ when coping with low energy processes. The new physics information parameterized in SMEFT is then inherited by LEFT through the matching conditions and renormalization group effects. Previously, the tree-level matching has been done in~\cite{Jenkins:2017jig} from the SMEFT effective interactions up to dim-6 operators to the LEFT also up to dim-6 operators. In this section we extend this matching to the dim-7 operators in both SMEFT and LEFT based on the basis of dim-7 operators in LEFT described in section~\ref{sec2} and the basis of dim-7 operators in SMEFT established in Ref.~\cite{Liao:2016hru} and further refined in Ref.~\cite{Liao:2020roy}. This result will be necessary for a consistent study of new physics effects at low energy beyond the leading order.

The matching is done by integrating out the SM heavy particles $W^\pm,~Z,~h,~t$ from the SMEFT in the electroweak symmetry broken phase. Since the effective interactions of higher-dimensional operators in SMEFT are supposed to be suppressed by more powers of $\LNP$ which is much larger than $\LEW$, we will work to the linear terms in them. Then the effective interaction of a dim-$m$ ($m\geq 5$) operator in SMEFT will possibly induce an effective interaction in LEFT of a dim-$n$ operator with the correspondence of the Wilson coefficients:
\begin{eqnarray}
\textrm{SMEFT: }C_{\rm SMEFT}^{{\rm dim-}m}\sim {1\over\Lambda_{\rm NP}^{m-4}} &\Rightarrow&
\textrm{LEFT: }C_{\rm LEFT}^{{\rm dim}-n}
\sim {1\over \Lambda_{\rm NP}^{m-4} \Lambda_{\rm EW}^{n-m}},
\end{eqnarray}
where we do not include couplings in SM. Since $t$ couples to another heavy particle ($W^\pm$) or another heavy particle ($Z,~h$) and itself, it cannot contribute to the tree-level matching up to dimension 7. Excluding the heavy particles ($W^\pm,~Z,~t$), $h$ couples very weakly to the light fermions. We will therefore ignore these small Yukawa couplings, so that the Higgs doublet field $H$ can be simply replaced by its vacuum expectation value (vev) $v/\sqrt{2}$ for the purpose of matching calculation. This leaves with us only the integration of the weak gauge bosons $W^\pm,~Z$. Inspection of the effective interactions from the dim-6 and dim-7 operators in SMEFT shows that a single $W^\pm,~Z$ propagator is required to connect an SMEFT vertex to an SM vertex to arrive at an LEFT operator up to dim-7.

We adopt for the dim-6 operators in SMEFT the Warsaw basis~\cite{Grzadkowski:2010es}, and for the dim-7 operators the basis in Ref.~\cite{Liao:2020roy} that is refined from the previous one~\cite{Liao:2016hru} and reproduced in table~\ref{tab3}. The bases of dim-5 and dim-6 operators in LEFT are taken from Ref.~\cite{Jenkins:2017jig} while the basis of dim-7 operators is listed in tables~\ref{tab1} and \ref{tab2}. Our matching results are recorded as follows. While the matching to dim-7 operators in LEFT is new, the matching results up to dim-6 operators in LEFT are to be added to those in Ref.~\cite{Jenkins:2017jig} when both baryon and lepton numbers match.

\begin{table}
\center
\scalebox{0.95}{
\begin{tabular}{|c|c|c|c|}
%\hline
 \multicolumn{2}{c}{$\psi^2H^4$} &  \multicolumn{2}{c}{ $\psi^2H^3D$}
\\
\hline
$\mathcal{O}_{LH}$ & $\epsilon_{ij}\epsilon_{mn}(\overline{L^{C,i}}L^m)H^jH^n(H^\dagger H)$ & 
$\mathcal{O}_{LeHD}$ & $\epsilon_{ij}\epsilon_{mn}(\overline{L^{C,i}}\gamma_\mu e)H^j(H^miD^\mu H^n)$
\\
\hline
\multicolumn{2}{c}{$\psi^2H^2D^2$}&  \multicolumn{2}{c}{$\psi^2H^2X$}
\\
\hline
$\mathcal{O}_{LDH1}$ & $\epsilon_{ij}\epsilon_{mn}(\overline{L^{C,i}}\overleftrightarrow{D_\mu} L^j)(H^mD^\mu H^n)$ &
$\mathcal{O}_{LHB}$ & $ g_1\epsilon_{ij}\epsilon_{mn}(\overline{L^{C,i}}
\sigma_{\mu\nu}L^m)H^jH^nB^{\mu\nu}$
\\
$\mathcal{O}_{LDH2}$ & $\epsilon_{im}\epsilon_{jn}(\overline{L^{C,i}}L^j)(D_\mu H^m D^\mu  H^n)$ &
$\mathcal{O}_{LHW}$ & $g_2\epsilon_{ij}(\epsilon \tau^I)_{mn}(\overline{L^{C,i}}\sigma_{\mu\nu}L^m)H^jH^nW^{I\mu\nu}$
\\
\hline
\multicolumn{2}{c}{$\psi^4D$}  &   \multicolumn{2}{c}{$\psi^4H$}
\\
\hline
$\mathcal{O}_{\overline{d}uLDL}$ & $\epsilon_{ij}(\overline{d}\gamma_\mu u)(\overline{L^{C,i}}i\overleftrightarrow{D}^\mu L^j)$ &
$\mathcal{O}_{\overline{e}LLLH}$ & $\epsilon_{ij}\epsilon_{mn}(\overline{e}L^i)(\overline{L^{C,j}}L^m)H^n$
\\
&  &
$\mathcal{O}_{\overline{d}QLLH1}$ & $\epsilon_{ij}\epsilon_{mn}(\overline{d}Q^i)(\overline{L^{C,j}}L^m)H^n$
\\
&  &
$\mathcal{O}_{\overline{d}QLLH2}$ & $\epsilon_{ij}\epsilon_{mn}(\overline{d}\sigma_{\mu\nu}Q^i)
(\overline{L^{C,j}}\sigma^{\mu\nu} L^m)H^n$
\\
&  &
$\mathcal{O}_{\overline{d}uLeH}$ & $\epsilon_{ij}(\overline{d}\gamma_\mu u)(\overline{L^{C,i}}\gamma^\mu e)H^j$
\\
&  &
$\mathcal{O}_{\overline{Q}uLLH}$ & $\epsilon_{ij}(\overline{Q}u)(\overline{L^{C}}L^i)H^j$
\\
\hline
\cellcolor{gray!25}$\mathcal{O}_{\overline{L}QdDd}$ &\cellcolor{gray!25} $(\overline{L}\gamma_\mu Q)(\overline{d^{C}}i\overleftrightarrow{D}^\mu d)$ &\cellcolor{gray!25}
$\mathcal{O}_{\overline{L}dud\tilde H}$ &\cellcolor{gray!25} $(\overline{L}d)(\overline{u^{C}}d)\tilde H$
\\
\cellcolor{gray!25}$\calO_{\overline{e}dddD}$  &\cellcolor{gray!25} $(\overline{e}\gamma_\mu d)(\overline{d^{C}}i\overleftrightarrow{D_\mu} d)$ &\cellcolor{gray!25}
$\calO_{\overline{L}dddH}$ &\cellcolor{gray!25} $(\overline{L}d)(\overline{d^{C}}d)H$
\\
\cellcolor{gray!25}&\cellcolor{gray!25}  &\cellcolor{gray!25}
$\calO_{\overline{e}Qdd\tilde H}$ &\cellcolor{gray!25} $\epsilon_{ij}(\overline{e}Q^{i})(\overline{d^{C}}d)\tilde H^j$
\\
\cellcolor{gray!25}&\cellcolor{gray!25}  &\cellcolor{gray!25}
$\calO_{\overline{L}dQQ\tilde H}$ &\cellcolor{gray!25} $\epsilon_{ij}(\overline{L}d)(\overline{Q^{C}}Q^{i})\tilde H^j$
\\
\hline
\end{tabular}
}
\caption{Basis of dim-7 operators in SMEFT~\cite{Liao:2020roy}. Here $L,~Q$ are the left-handed lepton and quark doublet fields, $u,~d,~e$ the right-handed up-type quark, down-type quark and charged lepton singlet fields, and $H$ the Higgs doublet with $\tilde H^i=\epsilon^{ij}H^{*j}$. Color contraction is implied for triple quark fields. The operators in gray have $(\Delta L,\Delta B)=(-1,1)$ while others have $(\Delta L,\Delta B)=(2,0)$. Hermitian conjugate operators are not displayed.}
\label{tab3}
\end{table}

%%%%%%%%%%%%%%%%%
\noindent
{\small$\scriptscriptstyle\blacksquare$ \small\bf\boldmath Matching from dim-5/7 operators in SMEFT to dim-3 operators in LEFT}
%%%%%%%%%%%%%%%%%
\begin{align}
\calO_\nu^{pr}=&(\overline{\nu^C_p}\nu_r), &
C_\nu^{pr}=&+\frac{1}{2}C_{5}^{ pr}v^2+\frac{1}{4}C_{LH}^{pr}v^4,
\label{eq_dim3}
\end{align}
where $C_5^{pr}$ is the Wilson coefficient of the dim-5 Weinberg operator $\calO_5=\epsilon_{ij}\epsilon_{mn}(\overline{L^{C,i}}L^m)H^jH^n$.

%%%%%%%%%%%%%%%%%
\noindent
{\small$\scriptscriptstyle\blacksquare$ \small\bf\boldmath Matching from dim-7 operators in SMEFT to dim-5 operators in LEFT}
%%%%%%%%%%%%%%%%%
\begin{align}
\calO_{\nu \gamma}^{pr}=&
(\overline{\nu^C_p}\sigma_{\mu\nu}\nu_r)F^{\mu\nu}, &
C_{\nu \gamma}^{pr}=&
+\frac{1}{4}ev^2\Big(2C_{LHB}^{pr}+C_{LHW}^{rp}-C_{LHW}^{pr}\Big),
\label{eq_dim5}
\end{align}
where the dim-5 Majorana neutrino dipole moment operator vanishes for identical flavors.

%%%%%%%%%%%%%%%%%
\noindent
{\small$\scriptscriptstyle\blacksquare$ \small\bf\boldmath Matching from dim-7 operators in SMEFT to dim-6 operators in LEFT}
%%%%%%%%%%%%%%%%%
\begin{itemize}
%%%
\item Operators with $(\Delta L,\Delta B)=(2,0)$:
\begin{align}
\nonumber
&\calO_{e\nu1}^{S,prst}=(\overline{e_{Rp}}e_{Lr})(\overline{\nu^C_s}\nu_t), &
&C_{e\nu1}^{S,prst}=-{\sqrt{2}v \over 8}\big(2C_{\bar{e}LLLH}^{prst}+C_{\bar{e}LLLH}^{psrt}+s\leftrightarrow t\big),
\\
\nonumber
&\calO_{e\nu2}^{S,prst}=(\overline{e_{Lp}}e_{Rr})(\overline{\nu^C_s}\nu_t), &
&C_{e\nu2}^{S,prst}=-{\sqrt{2}v \over 2}\big( C_{LeHD}^{sr}\delta^{tp}+ C_{LeHD}^{tr}\delta^{sp} \big),
\\
\nonumber
&\calO_{e\nu}^{T,prst}=(\overline{e_{Rp}}\sigma_{\mu\nu}e_{Lr})
(\overline{\nu^C_s}\sigma^{\mu\nu}\nu_t), &
&C_{e\nu}^{T}=+{\sqrt{2}v \over 32}\big( C_{\bar{e}LLLH}^{psrt}-C_{\bar{e}LLLH}^{ptrs} \big),
\\
\nonumber
&\calO_{d\nu}^{S,prst}=(\overline{d_{Rp}}d_{Lr})(\overline{\nu^C_s}\nu_t), &
&C_{d\nu}^{S,prst}=-{\sqrt{2}v \over 4}V_{xr}\big(C_{\bar dQLLH1}^{pxst}+C_{\bar dQLLH1}^{pxts}\big),
\\
\nonumber
&\calO_{d\nu}^{T,prst}=(\overline{d_{Rp}}\sigma_{\mu\nu}d_{Lr})
(\overline{\nu^C_s}\sigma^{\mu\nu}\nu_t), &
&C_{d\nu}^{T,prst}=-{\sqrt{2}v \over 4}V_{xr}\big( C_{\bar dQLLH2}^{pxst}-C_{\bar dQLLH2}^{pxts}\big),
\\
\nonumber
&\calO_{u\nu}^{S,prst}=(\overline{u_{Lp}}u_{Rr})(\overline{\nu^C_s}\nu_t), &
&C_{u\nu}^{S,prst}=+{\sqrt{2}v \over 4}\big( C_{\bar QuLLH}^{prst}+C_{\bar QuLLH}^{prts}\big),
\\
\nonumber
&\calO_{du\nu\,e1}^{S,prst}=(\overline{d_{Rp}}u_{Lr})
(\overline{\nu^C_s}e_{Lt}), &
&C_{du\nu\,e1}^{S,prst}=+{\sqrt{2}v \over 2}C_{\bar dQLLH1}^{prts},
\\
\nonumber
&\calO_{du\nu\,e2}^{S,prst}=(\overline{d_{Lp}}u_{Rr})
(\overline{\nu^C_s}e_{Lt}), &
&C_{du\nu\,e2}^{S,prst}=+{\sqrt{2}v \over 2}V_{xp}^*C_{\bar QuLLH}^{xrts},
\\
\nonumber
&\calO_{du\nu\,e}^{T,prst}=(\overline{d_{Rp}}\sigma_{\mu\nu}u_{Lr})
(\overline{\nu^C_s}\sigma^{\mu\nu}e_{Lt}), &
&C_{du\nu\,e}^{T,prst}=-{\sqrt{2}v \over 2}C_{\bar dQLLH2}^{prts},
\\
\nonumber
&\calO_{du\nu\,e1}^{V,prst}=(\overline{d_{Lp}}\gamma_\mu u_{Lr})(\overline{\nu^C_s}\gamma^\mu e_{Rt}), &
&C_{du\nu\,e1}^{V,prst}=+{ \sqrt{2}v \over 2 }V_{rp}^*C_{LeHD}^{st},
\\
&\calO_{du\nu\,e2}^{V,prst}=(\overline{d_{Rp}}\gamma_\mu u_{Rr})(\overline{\nu^C_s}\gamma^\mu e_{Rt}), &
&C_{du\nu\,e2}^{V,prst}=+{ \sqrt{2}v \over 2 }C_{\bar duLeH}^{prst},
\label{eq_dim6a}
\end{align}
where $V_{pr}$ is the CKM matrix coming from the SM charged current weak interactions. These matching results can contribute to nuclear neutrinoless double $\beta$ decays and LNV meson decays via the long distance mechanism~\cite{Liao:2019tep,Cirigliano:2017djv, Liao:2019gex,Liao:2020roy}.

\item Operators with $(\Delta L,\Delta B)=(-1,1)$:
\begin{align}
\nonumber
&\calO_{\nu\,dud1}^{S,prst}=
\epsilon_{\alpha\beta\gamma}(\overline{\nu_p}d_{Rr}^\alpha)
(\overline{u^{\beta C}_{Rs}}d_{Rt}^\gamma), &
&C_{\nu\,dud1}^{S,prst}=+{ \sqrt{2}v \over 2 }C_{\bar{L}dud\tilde H}^{prst},
\\
\nonumber
&\calO_{\nu\,dud2}^{S,prst}=
\epsilon_{\alpha\beta\gamma}(\overline{\nu_p}d_{Rr}^\alpha)
(\overline{u^{\beta C}_{Ls}}d_{Lt}^\gamma), &
&C_{\nu\,dud2}^{S,prst}=-{\sqrt{2}v\over 2} V_{xt}C_{\bar{L}dQQ\tilde H}^{prsx},
\\
\nonumber
&\calO_{eddd1}^{S,prst}=
\epsilon_{\alpha\beta\gamma}(\overline{e_{Lp}}d_{Rr}^\alpha)
(\overline{d^{\beta C}_{Rs}}d_{Rt}^\gamma), &
&C_{eddd1}^{S,prst}=+{ \sqrt{2}v \over 2 }C_{\bar{L}dddH}^{prst},
\\
\nonumber
&\calO_{eddd2}^{S,prst}=
\epsilon_{\alpha\beta\gamma}(\overline{e_{Rp}}d_{Lr}^\alpha)
(\overline{d^{\beta C}_{Rs}}d_{Rt}^\gamma), &
&C_{eddd2}^{S,prst}=-{ \sqrt{2}v \over 2 } V_{xr}C_{\bar{e}Qdd\tilde H}^{pxst},
\\
&\calO_{eddd3}^{S,prst}=
\epsilon_{\alpha\beta\gamma}(\overline{e_{Lp}}d_{Rr}^\alpha)
(\overline{d^{\beta C}_{Ls}}d_{Lt}^\gamma), &
&C_{eddd3}^{S,prst}=-{ \sqrt{2}v \over 4 }V_{xs}V_{yt}
\big( C_{\bar{L}dQQ\tilde H}^{prxy}-C_{\bar{L}dQQ\tilde H}^{pryx} \big).
\label{eq_dim6b}
\end{align}
\end{itemize}
These operators can induce usual nucleon decays such as $p\to\nu\pi^+$~\cite{Liao:2016hru} and $n\to e\pi^+$ that change baryon and lepton numbers by one unit while keeping their sum conserved.

%%%%%%%%%%%%%%%%%
\noindent
{\small$\scriptscriptstyle\blacksquare$ \small\bf\boldmath Matching from dim-7 operators in SMEFT to dim-7 operators in LEFT}
%%%%%%%%%%%%%%%%%
\begin{itemize}
%%%
\item Operators with $(\Delta L,\Delta B)=(2,0)$:
\begin{align}\nonumber
&\calO_{\nu\nu D}^{prst}=(\overline{\nu_p}\gamma^\mu \nu_r)\left(\overline{\nu^C_s}i\overleftrightarrow{\partial_\mu} \nu_t\right), &
&C_{\nu\nu D}^{prst}= -\delta^{pr}C_{LX}^{st},
\\\nonumber
&\calO_{e\nu D1}^{prst}=(\overline{e_{Lp}}\gamma^\mu e_{Lr})\left(\overline{\nu^C_s}i\overleftrightarrow{\partial_\mu} \nu_t\right), &
&C_{e\nu D1}^{prst}=+2\left({1\over 2}-s_W^2\right)\delta^{pr}C_{LX}^{st}
\\\nonumber
& & &+\left[\delta^{pt}\left( 2C_{LHW}^{sr}+C_{LDH1}^{sr}\right) -s\leftrightarrow t\right],
\nonumber
\\\nonumber
&\calO_{e\nu D2}^{prst}=(\overline{e_{Rp}}\gamma^\mu e_{Rr})\left(\overline{\nu^C_s}i\overleftrightarrow{\partial_\mu} \nu_t\right), &
&C_{e\nu D2}^{prst}=-2s_W^2\delta^{pr}C_{LX}^{st}
\\\nonumber
&\calO_{d\nu D1}^{prst}=(\overline{d_{Lp}}\gamma^\mu d_{Lr})\left(\overline{\nu^C_s}i\overleftrightarrow{\partial_\mu} \nu_t\right), &
&C_{d\nu D1}^{prst}=+2\left({1\over 2}-{1\over 3}s_W^2\right)\delta^{pr}C_{LX}^{st},
\\\nonumber
&\calO_{d\nu D2}^{prst}=(\overline{d_{Rp}}\gamma^\mu d_{Rr})\left(\overline{\nu^C_s}i\overleftrightarrow{\partial_\mu} \nu_t\right), &
&C_{d\nu D2}^{prst}=-{2\over 3}s_W^2\delta^{pr}C_{LX}^{st},
\\\nonumber
&\calO_{u\nu D1}^{prst}=(\overline{u_{Lp}}\gamma^\mu u_{Lr})\left(\overline{\nu^C_s}i\overleftrightarrow{\partial_\mu} \nu_t\right), &
&C_{u\nu D1}^{prst}=-2\left({1\over 2}-{2\over 3}s_W^2\right)\delta^{pr}C_{LX}^{st},
\\\nonumber
&\calO_{u D2}^{prst}=(\overline{u_{Rp}}\gamma^\mu u_{Rr})\left(\overline{\nu^C_s}i\overleftrightarrow{\partial_\mu} \nu_t\right), &
&C_{\nu\nu D}^{prst}=+{4\over 3}s_W^2\delta^{pr}C_{LX}^{st},
\\\nonumber
&\calO_{du\nu eD1}^{prst}=(\overline{d_{Lp}}\gamma^\mu u_{Lr})\left(\overline{e_{Ls}^C}i\overleftrightarrow{D_\mu} \nu_t\right), &
&C_{due\nu D1}^{prst}=+2V_{rp}^*\big( 2C_{LHW}^{ts}+C_{LDH1}^{ts}\big),
\\
&\calO_{due\nu D2}^{prst}=(\overline{d_{Rp}}\gamma^\mu u_{Rr})\left(\overline{e_{Ls}^C}i\overleftrightarrow{D_\mu} \nu_t\right), &
&C_{due\nu D2}^{prst}=-2C_{\bar{d}uLDL}^{prts},
\end{align}
where $s_W=\sin\theta_W,~c_W=\cos\theta_W$ with $\theta_W$ being the weak mixing angle, and the following shortcut is used,
\begin{align}
C_{LX}^{st}=2s_W^2C_{LHB}^{st} +c_W^2(C_{LHW}^{st}-C_{LHW}^{ts}).
\end{align}

%%%
\item Operators with $(\Delta L,\Delta B)=(1,-1)$:
\begin{align}
\nonumber
\calO_{u\nu dD2}^{prst}=&
\epsilon_{\alpha\beta\gamma}(\overline{u_L^\alpha}\gamma^\mu \nu)
(\overline{d_R^\beta}i\overleftrightarrow{D_\mu}d_R^{\gamma C}),
& C_{u\nu dD2}^{prst}=&-C_{\bar LQdDd}^{rpst*},
\\
\nonumber
\calO_{dedD2}^{prst}=&
\epsilon_{\alpha\beta\gamma}(\overline{d_L^\alpha}\gamma^\mu e_L)
(\overline{d_R^\beta}i\overleftrightarrow{D_\mu} d_R^{\gamma C}),
& C_{dedD2}^{prst}=&-V_{xp}^*C_{\bar LQdDd}^{rxst*},
\\
\calO_{dedD4}^{prst}=&
\epsilon_{\alpha\beta\gamma}(\overline{d_R^\alpha}\gamma^\mu e_R)
(\overline{d_R^\beta}i\overleftrightarrow{D_\mu} d_R^{\gamma C}),
& C_{dedD4}^{prst}=&-C_{\bar edddD}^{rpst*}.
\end{align}
\end{itemize}

%%%%%%%%%%%%%%%%%
\noindent
{\small$\scriptscriptstyle\blacksquare$ \small\bf\boldmath Matching from dim-6 operators in SMEFT to dim-7 operators in LEFT}
%%%%%%%%%%%%%%%%%

The operators involved in this matching all conserve baryon and lepton numbers. We will use the shortcuts:
\begin{align}
C_{eX}^{st}=c_W C_{eW}^{st}+s_W C_{eB}^{st},~~
C_{dX}^{st}=c_W C_{dW}^{st}+ s_W C_{dB}^{st},~~
C_{uX}^{st}=c_W C_{uW}^{st} - s_W C_{uB}^{st}.
\end{align}
\begin{itemize}
\item Operators in the class $(\bar{L}\gamma^\mu L)(\bar{L}iD_\mu R)$:
\begin{align}
\nonumber
&\calO_{\nu eD}^{prst}=(\overline{\nu_p}\gamma^\mu \nu_r)
(\overline{e_{Ls}} i\overleftrightarrow{D_\mu} e_{Rt}) , &
&C_{\nu eD}^{prst}
=-\frac{\sqrt{2}}{m_Z}\delta^{pr}C_{eX}^{st}
-\frac{2\sqrt{2}}{m_W}\delta^{sr}C_{eW}^{pt},
\\
\nonumber
&\calO_{\nu dD}^{prst}=(\overline{\nu_p}\gamma^\mu \nu_r)
(\overline{d_{Ls}} i\overleftrightarrow{D_\mu} d_{Rt}) , &
&C_{\nu dD}^{prst}=
-\frac{\sqrt{2}}{m_Z}\delta^{pr}V^*_{xs}C_{dX}^{xt},
\\
\nonumber
&\calO_{\nu uD}^{prst}=(\overline{\nu_p}\gamma^\mu \nu_r)
(\overline{u_{Ls}} i\overleftrightarrow{D_\mu} u_{Rt}) , &
&C_{\nu uD}^{prst}=
+\frac{\sqrt{2}}{m_Z}\delta^{pr}C_{uX}^{st},
\\
\nonumber
&\calO_{e eD1}^{prst}=(\overline{e_{Lp}}\gamma^\mu e_{Lr})
(\overline{e_{Ls}} i\overleftrightarrow{D_\mu} e_{Rt}),  &
&C_{e eD1}^{prst}=+\frac{2\sqrt{2}}{m_Z}
\left({1\over2}-s_W^2\right)\delta^{pr}C_{eX}^{st},
\\
\nonumber
&\calO_{edD1}^{prst}=(\overline{e_{Lp}}\gamma^\mu e_{Lr})
(\overline{d_{Ls}} i\overleftrightarrow{D_\mu} d_{Rt}),  &
&C_{edD1}^{prst}=+\frac{2\sqrt{2}}{m_Z}
\left({1\over2}-s_W^2\right)\delta^{pr}V^*_{xs}C_{dX}^{xt},
\\
\nonumber
&\calO_{euD1}^{prst}=(\overline{e_{Lp}}\gamma^\mu e_{Lr})
(\overline{u_{Ls}} i\overleftrightarrow{D_\mu} u_{Rt}),  &
&C_{euD1}^{prst}=-\frac{2\sqrt{2}}{m_Z}
\left({1\over2}-s_W^2\right)\delta^{pr}C_{uX}^{st},
\\
\nonumber
&\calO_{d eD1}^{prst}=(\overline{d_{Lp}}\gamma^\mu d_{Lr})
(\overline{e_{Ls}} i\overleftrightarrow{D_\mu} e_{Rt}), &
&C_{d eD1}^{prst}=+\frac{2\sqrt{2}}{m_Z}
\left({1\over2}-{1\over3}s_W^2\right)\delta^{pr}C_{eX}^{st},
\\
\nonumber
&\calO_{d dD1}^{prst}=(\overline{d_{Lp}}\gamma^\mu d_{Lr})
(\overline{d_{Ls}} i\overleftrightarrow{D_\mu} d_{Rt}),  &
&C_{d dD1}^{prst}=+\frac{2\sqrt{2}}{m_Z}
\left({1\over2}-{1\over3}s_W^2\right)\delta^{pr}V^*_{xs}C_{dX}^{xt},
\\
\nonumber
&\calO_{d uD1}^{prst}=(\overline{d_{Lp}}\gamma^\mu d_{Lr})
(\overline{u_{Ls}} i\overleftrightarrow{D_\mu} u_{Rt}) , &
&C_{d uD1}^{prst}=-\frac{2\sqrt{2}}{m_Z}
\left({1\over2}-{1\over3}s_W^2\right)\delta^{pr}C_{uX}^{st},
\\
\nonumber
&\calO_{ duD2}^{prst}=(\overline{d_{Lp}}\gamma^\mu d_{Lr}][\overline{u_{Ls}} i\overleftrightarrow{D_\mu} u_{Rt}) , &
&C_{ duD2}^{prst}=-\frac{2\sqrt{2}}{m_W}V_{sr}V^*_{xp}C_{uW}^{xt},
\\
\nonumber
&\calO_{u eD1}^{prst}=(\overline{u_{Lp}}\gamma^\mu u_{Lr})
(\overline{e_{Ls}} i\overleftrightarrow{D_\mu} e_{Rt}), &
&C_{u eD1}^{prst}=-\frac{2\sqrt{2}}{m_Z}
\left({1\over2}-{2\over3}s_W^2\right)\delta^{pr}C_{eX}^{st},
\\
\nonumber
&\calO_{u dD1}^{prst}=(\overline{u_{Lp}}\gamma^\mu u_{Lr})
(\overline{d_{Ls}} i\overleftrightarrow{D_\mu} d_{Rt}), &
&C_{u dD1}^{prst}=-\frac{2\sqrt{2}}{m_Z}
\left({1\over2}-{2\over3}s_W^2\right)\delta^{pr}V^*_{xs}C_{dX}^{xt},
\\
\nonumber
&\calO_{udD2}^{prst}=(\overline{u_{Lp}}\gamma^\mu u_{Lr}][\overline{d_{Ls}} i\overleftrightarrow{D_\mu} d_{Rt}), &
&C_{udD2}^{prst}=-\frac{2\sqrt{2}}{m_W}V^*_{rs}C_{dW}^{pt },
\\
\nonumber
&\calO_{u uD1}^{prst}=(\overline{u_{Lp}}\gamma^\mu u_{Lr})
(\overline{u_{Ls}} i\overleftrightarrow{D_\mu} u_{Rt}), &
&C_{u uD1}^{prst}=+\frac{2\sqrt{2}}{m_Z}\delta^{pr}
\left({1\over2}-{2\over3}s_W^2\right)C_{uX}^{st},
\\
\nonumber
&\calO_{\nu eduD}^{prst}=(\overline{\nu_p}\gamma^\mu e_{Lr})(\overline{d_{Ls}} i\overleftrightarrow{D_\mu} u_{Rt}), &
&C_{\nu eduD}^{prst}=+\frac{2\sqrt{2}}{m_W}\delta^{pr} V^*_{ws}C_{uW}^{wt},
\\
\nonumber
&\calO_{e\nu udD}^{prst}=(\overline{e_{Lp}}\gamma^\mu \nu_r)(\overline{u_{Ls}} i\overleftrightarrow{D_\mu} d_{Rt}), &
&C_{e\nu udD}^{prst}=+\frac{2\sqrt{2}}{m_W}\delta^{pr} C_{dW}^{st},
\\
&\calO_{du \nu eD1}^{prst}=(\overline{d_{Lp}}\gamma^\mu u_{Lr})(\overline{\nu_s} i\overleftrightarrow{D_\mu} e_{Rt}),&
&C_{du \nu eD1}^{prst}=+\frac{2\sqrt{2}}{m_W}V^*_{rp}C_{eW}^{st }.
\label{eq_dim7a}
\end{align}
%%%%%%%%
\item Operators in the class $(\bar{R}\gamma^\mu R)(\bar{L}iD_\mu R)$:
\begin{align}\nonumber
\calO_{e eD2}^{prst}=&(\overline{e_{Rp}}\gamma^\mu e_{Rr})(\overline{e_{Ls}} i\overleftrightarrow{D_\mu} e_{Rt}), &
C_{e eD2}^{prst}=&
-\frac{2\sqrt{2}}{m_Z}s_W^2\delta^{pr}C_{eX}^{st},
\\
\nonumber
\calO_{edD2}^{prst}=&(\overline{e_{Rp}}\gamma^\mu e_{Rr})(\overline{d_{Ls}} i\overleftrightarrow{D_\mu} d_{Rt}), &
C_{edD2}^{prst}=&
-\frac{2\sqrt{2}}{m_Z}s_W^2\delta^{pr}V^*_{xs}C_{dX}^{xt},
\\
\nonumber
\calO_{euD2}^{prst}=&(\overline{e_{Rp}}\gamma^\mu e_{Rr})(\overline{u_{Ls}} i\overleftrightarrow{D_\mu} u_{Rt}), &
C_{euD2}^{prst}=&
+\frac{2\sqrt{2}}{m_Z}s_W^2\delta^{pr}C_{uX}^{st},
\\
\nonumber
\calO_{d eD2}^{prst}=&(\overline{d_{Rp}}\gamma^\mu d_{Rr})(\overline{e_{Ls}} i\overleftrightarrow{D_\mu} e_{Rt}), &
C_{d eD2}^{prst}=&
-\frac{2\sqrt{2}}{m_Z}{1\over 3}s_W^2\delta^{pr}C_{eX}^{st},
\\
\nonumber
\calO_{d dD2}^{prst}=&(\overline{d_{Rp}}\gamma^\mu d_{Rr})(\overline{d_{Ls}} i\overleftrightarrow{D_\mu} d_{Rt}), &
C_{d dD2}^{prst}=&
-\frac{2\sqrt{2}}{m_Z}{1\over 3}s_W^2\delta^{pr}V^*_{xs}C_{dX}^{xt},
\\
\nonumber
\calO_{d uD3}^{prst}=&(\overline{d_{Rp}}\gamma^\mu d_{Rr})(\overline{u_{Ls}} i\overleftrightarrow{D_\mu} u_{Rt}), &
C_{d uD3}^{prst}=&
+\frac{2\sqrt{2}}{m_Z}{1\over 3}s_W^2\delta^{pr}C_{uX}^{st},
\\
\nonumber
\calO_{u eD2}^{prst}=&(\overline{u_{Rp}}\gamma^\mu u_{Rr})(\overline{e_{Ls}} i\overleftrightarrow{D_\mu} e_{Rt}), &
C_{u eD2}^{prst}=&
+\frac{2\sqrt{2}}{m_Z}{2\over 3}s_W^2\delta^{pr}C_{eX}^{st},
\\
\nonumber
\calO_{u dD3}^{prst}=&(\overline{u_{Rp}}\gamma^\mu u_{Rr})(\overline{d_{Ls}} i\overleftrightarrow{D_\mu} d_{Rt}), &
C_{u dD3}^{prst}=&
+\frac{2\sqrt{2}}{m_Z}{2\over 3}s_W^2\delta^{pr}V^*_{xs}C_{dX}^{xt},
\\
\calO_{u uD2}^{prst}=&(\overline{u_{Rp}}\gamma^\mu u_{Rr})(\overline{u_{Ls}} i\overleftrightarrow{D_\mu} u_{Rt}), &
C_{u uD2}^{prst}=&
-\frac{2\sqrt{2}}{m_Z}{2\over 3}s_W^2\delta^{pr}C_{uX}^{st}.
\label{eq_dim7b}
\end{align}
\end{itemize}

%%%%%%%%%%%%%%%%%%%%%%%
\section{Low energy neutrino-photon interactions and ultraviolet completion}
%%%%%%%%%%%%%%%%%%%%%%%
\label{sec4}

Among the dim-7 operators in LEFT the most interesting might be the $\psi^2X^2$ type, which contains both $(\Delta L,\Delta B)=(0,0)$ and  $(\Delta L,\Delta B)=(2,0)$ sectors. We first note that Ref.~\cite{Chalons:2013mya} listed the subset of dim-7 operators for the $b\rightarrow s$ transition which however contains an identically vanishing operator $\mathcal{E}^T_{L,R}=\bar{b}\sigma^{\nu\rho}P_{L/R}s
F_{\mu\nu}F^\mu_{~\rho}=0$. In this section we consider the low energy neutrino-photon ($\nu\gamma$) interactions in the $(\Delta L,\Delta B)=(2,0)$ sector. The leading terms appear at dimension 7:
\begin{eqnarray}
{\cal L}_{\nu\gamma}^\textrm{LNV}&=&
\calO_{\nu F1}^{\alpha\beta}C_{\nu F1}^{\alpha\beta}
+\calO_{\nu F2}^{\alpha\beta}C_{\nu F2}^{\alpha\beta}+{\rm h.c.},
\label{eq_nuphoton}
\end{eqnarray}
where the two operators are listed in table~\ref{tab2} whose Wilson coefficients are symmetric in neutrino flavors $\alpha,~\beta$. We note in passing that in Ref.~\cite{Altmannshofer:2018xyo} the operators $\calO_{\nu F1/2}$ and $\calO_{\nu G1/2}$ have been used to study coherent elastic neutrino-nucleus scattering. As a neutral particle, these interactions cannot originate directly from a tree level matching to the first few high-dimensional operators in SMEFT. Instead, they would arise as a loop effect of the effective interactions between neutrinos and charged particles in LEFT that can originate from a tree level matching to SMEFT. We content ourselves in this work with effective $\nu\gamma$ interactions at energies below the mass of the lightest charged particle, i.e., the electron. We will see that the dominant contribution comes from the dim-6 operators~\cite{Jenkins:2017jig} involving the electron:
\begin{eqnarray}
\calL_{\nu e}^6&=&
\calO_{e\nu1}^{S,ee\alpha\beta}C_{e\nu1}^{S,ee\alpha\beta}
+\calO_{e\nu2}^{S,ee\alpha\beta}C_{e\nu2}^{S,ee\alpha\beta}
+\calO_{e\nu}^{T,ee\alpha\beta}C_{e\nu}^{T,ee\alpha\beta}+{\rm h.c.},
\label{eq_nue}
\end{eqnarray}
where the Wilson coefficients are given in equation~\eqref{eq_dim6a} in terms of those in SMEFT.

\begin{figure}
\centering
\includegraphics[width=13cm]{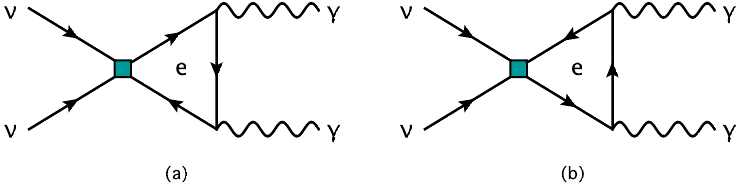}
\caption{One-loop Feynman diagrams that induce effective $\nu\gamma$ interactions in equation~\eqref{eq_nuphoton} due to effective $\nu e$ interactions in equation~\eqref{eq_nue}.}
\label{fig1}
\end{figure}

Contracting the two electron lines in any of the vertices in equation~\eqref{eq_nue} and attaching two photons to the contracted electron line yields the effective interactions between two neutrinos and two photons as shown in figure~\ref{fig1}, which at energies below the electron mass $m_e$ have the form of equation~\eqref{eq_nuphoton}, with
\begin{equation}
C_{\nu F1}^{\alpha\beta}={1 \over 12\pi m_e} \left(C_{e\nu1}^{S,ee\alpha\beta}+C_{e\nu2}^{S,ee\alpha\beta}\right),~~~
C_{\nu F2}^{\alpha\beta}=-{i \over 8\pi m_e} \left(C_{e\nu1}^{S,ee\alpha\beta}-C_{e\nu2}^{S,ee\alpha\beta}\right).
\label{eq_CnuF}
\end{equation}
The tensor interaction in equation~\eqref{eq_nue} yields a vanishing result because of the Schouten identity,
\begin{equation}
g^{\alpha\beta}\epsilon^{\mu\nu\rho\sigma}
+g^{\alpha\mu}\epsilon^{\nu\rho\sigma\beta}
+g^{\alpha\nu}\epsilon^{\rho\sigma\beta\mu}
+g^{\alpha\rho}\epsilon^{\sigma\beta\mu\nu}
+g^{\alpha\sigma}\epsilon^{\beta\mu\nu\rho}=0,
\end{equation}
which is indeed consistent with the absence in table~\ref{tab2} of a neutrino tensor bilinear coupled to a field strength squared. The $1/m_e$ factor in equation~\eqref{eq_CnuF} is not surprising, but is actually the same as that in the $t$-loop contribution to the decay amplitude for $h\to\gamma\gamma$ in the heavy top limit where $1/m_t$ is cancelled by the top Yukawa coupling. There is an additional contribution to the Wilson coefficient $C_{\nu F1}$: when one $H$ in the dim-5 Weinberg operator $\calO_5$ assumes its vev and the other $H$ field is connected to the two photons through the SM one-loop diagrams, $C_{\nu F1}$ gains a term proportional to $m_\nu/(v^2m_h^2)$ which is suppressed by the neutrino mass $m_\nu$ and can be safely ignored. Parameterizing by $\LNP^{-3}$ the SMEFT Wilson coefficients entering $C_{e\nu1(2)}^{S,ee\alpha\beta}$ through the matching conditions in equation~\eqref{eq_dim6a}, one has roughly
\begin{equation}
C_{\nu F1(2)}^{\alpha\beta}\sim{v\over 4\pi m_e} \frac{1}{\LNP^3},
\end{equation}
which offers a huge enhancement factor of $\sim 10^4-10^5$ compared to the effect of a usual dim-7 operator.

%%%%%

With the above enhancement in mind we calculate the cross sections for various $\nu\gamma$ scattering processes. The amplitudes are
\begin{align}
\calA(\gamma_\lambda(k)\nu_\alpha(p)\to
\gamma_{\lambda^\prime}(k^\prime)\bar\nu_\beta(p^\prime))
&=2\alpha_{\rm em}\left[C_{\nu F1}^{\alpha\beta}\left(1-\lambda\lambda^\prime\right)
+iC_{\nu F2}^{\alpha\beta}\left(\lambda-\lambda^\prime\right)\right](-t)^{3/2},
\nonumber
\\
\calA(\gamma_\lambda(k)\gamma_{\lambda^\prime}(k^\prime)\to
\nu_\alpha(p)\nu_\beta(p^\prime))&=
2\alpha_{\rm em}\left[ C_{\nu F1}^{\alpha\beta*}(1+\lambda\lambda^\prime)
+iC_{\nu F2}^{\alpha\beta*}(\lambda+\lambda^\prime)\right]s^{3/2},
\nonumber
\\
\calA(\nu_\alpha(p)\nu_\beta(p^\prime)\to \gamma_\lambda(k)\gamma_{\lambda^\prime}(k^\prime))&=
2\alpha_{\rm em}\left[ C_{\nu F1}^{\alpha\beta}(1+\lambda\lambda^\prime)
+iC_{\nu F2}^{\alpha\beta}(\lambda+\lambda^\prime)\right]s^{3/2}.
\end{align}
Here $\lambda,~\lambda^\prime$ denote the helicities of the photons, $s=(k+k')^2$, and $t=(k-k')^2$. We have ignored the tiny masses of the neutrinos and explicitly evaluated their spinor wavefunctions. The crossing symmetry is manifest in the above amplitudes: the first and third amplitudes are related by $(s,\lambda')\leftrightarrow(-t,-\lambda')$ while the last two are related by $(\lambda,\lambda')\leftrightarrow(-\lambda,-\lambda')$ and complex conjugate. Denoting the photon energy by $\omega$ and the scattering angle by $\theta$ in the center of mass frame, the differential cross sections are,
\begin{align}
{d\sigma(\nu_\alpha\gamma_\lambda\to\bar\nu_\beta\gamma_{\lambda^\prime})
\over d\cos\theta}
=&{ \alpha_{\rm em}^2\omega^4\over 4\pi}
\left|C_{\nu F1}^{\alpha\beta} \left(1-\lambda\lambda^\prime\right)+iC_{\nu F2}^{\alpha\beta}\left(\lambda-\lambda^\prime\right)\right|^2
(1-\cos\theta)^3,
\nonumber
\\
{d\sigma(\gamma_\lambda\gamma_{\lambda^\prime}\to\nu_\alpha\nu_\beta)
\over d\cos\theta}=&
{\alpha_{\rm em}^2\omega^4 \over \pi } \left|C_{\nu F1}^{\alpha\beta}\left(1+\lambda\lambda^\prime\right)+iC_{\nu F2}^{\alpha\beta}\left(\lambda+\lambda^\prime\right)\right|^2
{2\over 1+\delta_{\alpha\beta}},
\nonumber
\\
{d\sigma(\nu_\alpha\nu_\beta\to\gamma_\lambda\gamma_{\lambda^\prime})
\over d\cos\theta}=&{ \alpha_{\rm em}^2\omega^4\over \pi }\left|C_{\nu F1}^{\alpha\beta}\left(1+\lambda\lambda^\prime\right)+iC_{\nu F2}^{\alpha\beta}\left(\lambda+\lambda^\prime\right)\right|^2.
\end{align}
Upon averaging (summing) over the initial (final) photon helicities, the total cross sections are,
\begin{eqnarray}
&&\sigma(\nu_\alpha\gamma\to\bar\nu_\beta\gamma)
=\frac{4\alpha_{\rm em}^2\omega^4}{\pi}\left(|C_{\nu F1}^{\alpha\beta}|^2+ |C_{\nu F2}^{\alpha\beta}|^2 \right),
\nonumber
\\
&&
\sigma(\gamma\gamma\to\nu_\alpha\nu_\beta)
=\sigma(\gamma\nu_\alpha\to\bar\nu_\beta\gamma)
{2\over 1+\delta_{\alpha\beta}},
\nonumber
\\
&&\sigma(\nu_\alpha\nu_\beta\to\gamma\gamma)
=4\sigma(\gamma\nu_\alpha\to\bar\nu_\beta\gamma).
\label{crosssection1}
\end{eqnarray}

There are some salient features in our above results when compared to their counterparts in SM~\cite{Dicus:1993iy}, i.e., $\nu\gamma\to\nu\gamma$, $\gamma\gamma\to\nu\bar\nu$, and $\nu\bar\nu\to\gamma\gamma$. First, the cross sections for each pair of similar processes vanish for opposite paring of photon helicities. For instance, $\nu\gamma_\lambda\to\bar\nu\gamma_{\lambda'}$ here does not occur for identical helicities $\lambda=\lambda'$, while $\nu\gamma_\lambda\to\nu\gamma_{\lambda'}$ in SM is absent for opposite helicities  $\lambda=-\lambda'$. The situation for the other two pairs of processes is just reversed. This circumstance is an interesting consequence of lepton number being violated or conserved: fixing an always left-handed neutrino in either initial or final states, what is for the second fermion to be a left-handed neutrino (right-handed antineutrino) in the SM process becomes a right-handed (left-handed) neutrino in the process under consideration here. Thus in a sense the flip or nonflip of a photon helicity offers a veto to Dirac or Majorana neutrinos. In addition, $\gamma\nu\to\gamma\bar\nu$ cannot take place in the forward direction, while $\gamma\gamma\to\nu\nu$ and $\nu\nu\to\gamma\gamma$ show a purely $s$-wave behavior. These are also different from the SM processes. Second, our cross sections are proportional to $(v^2/m_e^2)\LNP^{-6}$ while the SM ones are typical one-loop processes of order $m_W^{-8}\ln^2(m_W^2/m_e^2)$. This results in a different low energy behavior in cross sections: our processes behave as $\omega^4$ while the SM ones go as $\omega^6$. All of this power counting is indeed consistent with the fact that the effective operators for $\nu\gamma$ interactions start with dimension 7 here but with dimension 8 in SM. Numerically, in our case $\sigma(\gamma\nu_\alpha\to\bar\nu_\beta\gamma)/
\sigma(\gamma\gamma\to\nu_\alpha\nu_\beta)=1~(1/2)$ for $\nu_\alpha=\nu_\beta$ ($\nu_\alpha\ne\nu_\beta$), which is in contrast to the SM case $\sigma(\nu\gamma\to\nu\gamma)/\sigma(\gamma\gamma\to\nu\bar\nu)\sim 15$.

%%%%%%%
\begin{figure}
\centering
\includegraphics[width=10cm]{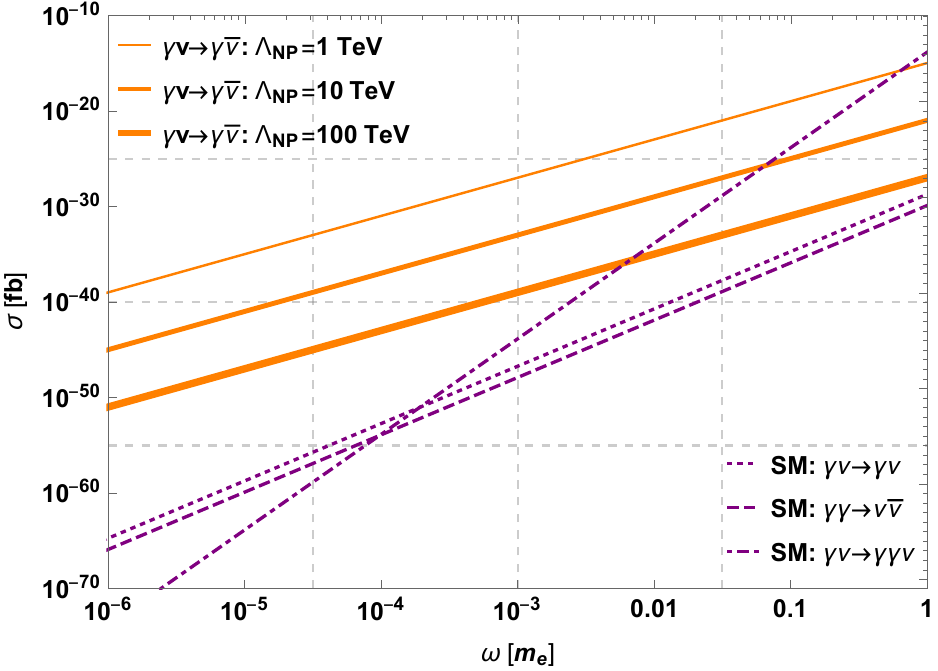}
\caption{Cross sections are shown as functions of photon energy $\omega$ (in units of $m_e$) for LNV scattering (orange solid curves) and SM scattering (purple dashed or dot-dashed curves)~\cite{Dicus:1993iy,Dicus:1997rw}.}
\label{fig2}
\end{figure}

To get some feel about the orders of magnitude of various processes here and in SM, we make a simplifying assumption in our matching conditions shown in equation~\eqref{eq_dim6a}, i.e., the Wilson coefficient $C_{LeHD}^{pr}=0$ in SMEFT, so that the Wilson coefficient $C_{e\nu 2}^{S,prst}=0$ in LEFT while $C_{e\nu 1}^{S,prst}$ gains a contribution from the Wilson coefficient $C_{\bar eLLLH}^{prst}$ in SMEFT. To compare with the SM processes, we consider the case with $\nu_\alpha=\nu_\beta\equiv\nu$, so that effectively we have from equations~\eqref{eq_dim6a} and \eqref{eq_CnuF},
\begin{equation}
C_{\nu F1}^{11}=-\frac{\sqrt{2}v}{16\pi m_e}C_{\bar{e}LLLH}^{1111},~~~
C_{\nu F2}^{11}=i\frac{3\sqrt{2}v}{32\pi m_e}C_{\bar{e}LLLH}^{1111},
\end{equation}
where the superscript 1 refers to the first generation neutrino and charged lepton. Parameterizing $|C_{\bar{e}LLLH}^{1111}|=\LNP^{-3}$, this gives
\begin{equation}
\sigma(\gamma\nu\to\bar\nu\gamma)
={13\alpha_{\rm em}^2 \over 128\pi^3}{v^2\over m_e^2}
{\omega^4\over \Lambda_{\rm NP}^6}
\approx 1.1\times10^{-15}\left({\omega\over m_e}\right)^4
\left({{\rm TeV} \over\Lambda_{\rm NP} }\right)^6\rm {fb},
\label{crosssection2}
\end{equation}
and $\sigma(\gamma\gamma\to\nu\nu)=\sigma(\gamma\nu\to\bar\nu\gamma)$,
$\sigma(\nu\nu\to\gamma\gamma)=4\sigma(\gamma\nu\to\bar\nu\gamma)$. The above cross section is depicted in figure~\ref{fig2} as a function of $\omega/m_e$ at three values of the new physics scale $\LNP=1,~10,~100~\TeV$. Also shown are the SM cross sections for $\gamma\nu\to\gamma\nu$, $\gamma\gamma\to\nu\bar\nu$~\cite{Dicus:1993iy}, and $\gamma\nu\to\gamma\gamma\nu$~\cite{Dicus:1997rw}. The last process arises from dim-10 operators whose Wilson coefficients are significantly enhanced at one loop by a factor of $1/m_e^4$, and has an $\omega^{10}$ behavior in its cross section. As one can see from the figure, the LNV $\nu\gamma$ interactions result in a generically much larger cross section even for a high scale $\LNP$ than the SM interactions. We will systematically explore in the future work its possible implications in cosmology.

\begin{figure}
\centering
\includegraphics[width=18cm]{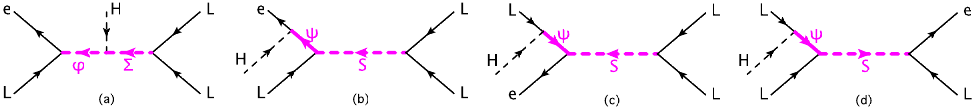}
\caption{Tree-level topologies for ultraviolet completion generating operator ${\cal O}_{\bar eLLLH}$.}
\label{fig3}
\end{figure}

Before we conclude this section we illustrate by examples how the dim-7 operators $\calO_{\bar eLLLH}^{prst}$ in SMEFT that are called for in the above analysis could be generated by ultraviolet completion. The possible tree-level topologies are classified in figure~\ref{fig3}. While the topology (a) only involves new scalar fields, the others require both scalar and fermion fields. We notice that gauge anomaly cancellation may demand vector-like fermions which are usually easy to arrange. The electroweak gauge symmetry $SU(2)_L\times U(1)_Y$ at each vertex then gives two possible solutions to the quantum numbers of the heavy fields in each topology, which are:
\begin{align}
\nonumber
&\textrm{model (a1)}:~\Sigma=(3,-1), ~~\varphi=\Big(2,-{1\over2}\Big), &
&\textrm{model (a2)}:~\Sigma=(1,-1), ~~\varphi=\Big(2,-{1\over2}\Big),
\\\nonumber
&\textrm{model (b1)}:~S=(3,-1), ~~\psi=\Big(2,-{3\over2}\Big), &
&\textrm{model (b2)}:~S=(1,-1), ~~\psi=\Big(2,-{3\over2}\Big),
\\\nonumber
&\textrm{model (c1)}:~S=(3,-1), ~~\psi=(3,0), &
&\textrm{model (c2)}:~S=(1,-1), ~~\psi=(1,0),
\\
&\textrm{model (d1)}:~S=\Big(2,-{1\over2}\Big), ~\psi=(3,0), &
&\textrm{model (d2)}:~S=\Big(2,-{1\over2}\Big), ~\psi=(1,0).
\end{align}

Let us consider model (a2) as an example. The relevant new terms in the Lagrangian are,
\begin{align}
{\cal L}_{\rm}\supset Y_{\Sigma,pr}\epsilon_{ij} \overline{L^{C,i}_p}L^j_r \Sigma^\dagger
+\lambda_{\Sigma\varphi} \Sigma \varphi^\dagger H+Y_{\varphi,pr}  \epsilon_{ij}\overline{e_p}L^i_r\varphi^j
+{\rm h.c.},
\end{align}
where $Y_{\Sigma}=-Y_{\Sigma}^{\rm T}$, $Y_{\varphi}$, and $\lambda_{\Sigma\varphi}$ are generally complex Yukawa coupling matrices in lepton flavors and triple scalar coupling respectively. Then the diagram (a) and its crossings in figure~\ref{fig3} lead to the effective interaction $C_{\bar eLLLH}^{prst}\calO_{\bar eLLLH}^{prst}$. But before we present the Wilson coefficients we must first decide on the set of independent operators contained in $\calO_{\bar eLLLH}^{prst}$ which have nontrivial flavor relations~\cite{Liao:2019tep}:
\begin{align}
{\cal O}_{\bar eLLLH}^{prst}+{\cal O}_{\bar eLLLH}^{ptsr}={\cal O}_{\bar eLLLH}^{psrt}+{\cal O}_{\bar eLLLH}^{ptrs}={\cal O}_{\bar eLLLH}^{pstr}+{\cal O}_{\bar eLLLH}^{prts}.
\end{align}
Note that the second equality is actually not independent but can be obtained from the first one, and we include it only for clarity. With three generations, suppose we choose the set to be,
\begin{equation}
{\cal O}_{\bar eLLLH}^{prrr},~{\cal O}_{\bar eLLLH}^{prss},~{\cal O}_{\bar eLLLH}^{pssr},~{\cal O}_{\bar eLLLH}^{p123},~{\cal O}_{\bar eLLLH}^{p132},~{\cal O}_{\bar eLLLH}^{p213},~{\cal O}_{\bar eLLLH}^{p231},
\label{eq_Oset}
\end{equation}
where $s\ne r$ assumes values $1,~2,~3$, then the redundant operators are,
\begin{align}
\calO^{p321}_{\overline{e}LLLH}&=
\calO^{p231}_{\overline{e}LLLH}
+\calO^{p132}_{\overline{e}LLLH}-\calO^{p123}_{\overline{e}LLLH},
\nonumber
\\
\calO^{p312}_{\overline{e}LLLH}&=
\calO^{p231}_{\overline{e}LLLH}
+\calO^{p132}_{\overline{e}LLLH}-\calO^{p213}_{\overline{e}LLLH},
\nonumber
\\
\calO^{prsr}_{\overline{e}LLLH}&=
{1\over 2}(\calO^{prss}_{\overline{e}LLLH}
+\calO^{pssr}_{\overline{e}LLLH}).
\label{eq_redundant}
\end{align}
By integrating out heavy particles from the Lagrangian or computing first amplitudes and then rewriting them back into effective interactions, we find the Wilson coefficients for the set of independent operators shown in equation~\eqref{eq_Oset} upon applying Fierz and other algebraic identities:
\begin{align}
C_{\bar eLLLH}^{prrr}&=0, &
C_{\bar eLLLH}^{prss}&=
-{\lambda_{\Sigma\varphi} \over m_\Sigma^2 m_\varphi^2}\calU_{ps;rs}
=-C_{\bar eLLLH}^{pssr},
\nonumber
\\
C_{\bar eLLLH}^{p123}&=
2{\lambda_{\Sigma\varphi} \over m_\Sigma^2 m_\varphi^2}
\big[\calU_{p1;32}-\calU_{p3;12}\big], &
C_{\bar eLLLH}^{p132}&=
2{\lambda_{\Sigma\varphi} \over m_\Sigma^2 m_\varphi^2}\calU_{p1;23},
\nonumber
\\
C_{\bar eLLLH}^{p213}&=
2{\lambda_{\Sigma\varphi} \over m_\Sigma^2 m_\varphi^2}
\big[ \calU_{p2;31}-\calU_{p3;21}\big], &
C_{\bar eLLLH}^{p231}&=
2{\lambda_{\Sigma\varphi} \over m_\Sigma^2 m_\varphi^2}\calU_{p2;13},
\end{align}
where $m_{\Sigma,\varphi}$ are the masses of the new heavy scalars and the shortcut $\calU_{pr;st}=(Y_\varphi)_{pr}(Y_\Sigma)_{st}$ is used. With the matching condition in equation~\eqref{eq_dim6a}, we obtain the corresponding LEFT Wilson coefficient in equation~\eqref{eq_nue}:
\begin{align}
 C_{e\nu1}^{S,ee\alpha\beta}
 =&{\sqrt{2}v\lambda_{\Sigma\varphi} \over 4m_\Sigma^2 m_\varphi^2}\left[ (Y_\varphi)_{1\alpha}(Y_\Sigma)_{1\beta}
 +(Y_\varphi)_{1\beta}(Y_\Sigma)_{1\alpha}\right].
\end{align}
It is interesting that while $C_{e\nu1}^{S,eeee}=0$ due to antisymmetry of the $Y_\Sigma$ matrix, $C_{e\nu1}^{S,ee\alpha\beta}$ generically does not vanish when either of the neutrino indices $\alpha,~\beta$ or both refers to the second or third generation.

As a second example we consider model (d2) which introduces three vector-like heavy singlet fermions $\psi$ of mass matrix $M_\psi$ and one doublet scalar $S$ of mass $m_S$. The relevant new Yukawa couplings are,
\begin{align}
\calL_{\rm yuk}=&
(Y_{H\psi})_{pr}\epsilon_{ij}\bar\psi_p L_r^i H^j
+(Y_{S\psi})_{pr}\delta_{ij}\overline{L^{C,i}_p}\psi_r S^{*j}
+(Y_{Se})_{pr}\epsilon_{ij}\overline{e_p}L_r^i S^j+{\rm h.c.},
\end{align}
where $Y_{H\psi},~Y_{S\psi},~Y_{Se}$ are complex Yukawa coupling matrices in generation space. Choosing the same set of independent operators in equation~\eqref{eq_Oset}, the diagram (d) in figure~\ref{fig3} leads to the tree-level result:
\begin{align}
&C_{\bar eLLLH}^{prrr}={1\over m_S^2}\calV_{pr;rr},
\nonumber
\\
&C_{\bar eLLLH}^{prss}={1\over m_S^2}
\Big[\calV_{pr;ss}+{1\over 2}\calV_{ps;rs}\Big], &
&C_{\bar eLLLH}^{pssr}={1\over m_S^2}
\Big[\calV_{ps;sr} +{1\over 2}\calV_{ps;rs}\Big],
\nonumber
\\
&C_{\bar eLLLH}^{p123}={1\over m_S^2}
\big[ \calV_{p1;23}-\calV_{p3;21}\big], &
&C_{\bar eLLLH}^{p132}={1\over m_S^2}
\big[\calV_{p1;32} +\calV_{p3;12}+\calV_{p3;21}\big],
\nonumber
\\
&C_{\bar eLLLH}^{p213}={1\over m_S^2}
\big[\calV_{p2;13}-\calV_{p3;12}\big], &
&C_{\bar eLLLH}^{p231}={1\over m_S^2}
\big[\calV_{p2;31} +\calV_{p3;12}+\calV_{p3;21}\big],
\end{align}
with the shortcut $\calV_{pr;st}=(Y_{Se})_{pr}(Y_{S\psi}M_\psi^{-1}Y_{H\psi})_{st}$.
Thus the LEFT Wilson coefficient in equation~\eqref{eq_nue} becomes
\begin{align}
C_{e\nu1}^{eeee}=-{3\sqrt{2}v\over 4m_S^2}(Y_{Se})_{11}
(Y_{S\psi}M_\psi^{-1}Y_{H\psi})_{11}.
\end{align}
We see that in this case $C_{e\nu1}^{S,eeee}$ with identical lepton flavors survives.

%%%%%%%%%%%%%%%%%%%%%%%
\section{Conclusion}
%%%%%%%%%%%%%%%%%%%%%%%
\label{sec5}

We have established the basis of dim-7 operators in LEFT which is a low energy effective field theory for the SM particles excluding the weak gauge bosons, the Higgs boson and the top quark. We found these operators are classified into four sectors according to their baryon and lepton numbers. Including Hermitian conjugates of the operators, there are 3168 operators with $(\Delta L,\Delta B)=(0,0)$, 750 operators with $(\Delta L,\Delta B)=(\pm 2,0)$, 588 operators with  $(\Delta L,\Delta B)=(\pm 1,\mp 1)$, and 712 operators with $(\Delta L,\Delta B)=(\pm 1,\pm 1)$. We have done a tree-level matching calculation to relate the Wilson coefficients between the SMEFT defined above the electroweak scale and the LEFT. The matching incorporates new terms due to dim-7 operators in SMEFT on the one hand, and extends to dim-7 operators in LEFT found in this work on the other.  As a phenomenological application we have calculated the effective neutrino-photon interaction due to dim-7 operators in LEFT, and found several interesting features compared to the SM case. The cross sections for neutrino-photon scattering have a different correlation between the helicities of the photons. The interaction arises from a one-loop effect due to dim-6 operators in LEFT and is significantly enhanced at low energy by an inverse electron mass. As a consequence of this, the cross sections are even larger than their SM counterparts for a new physics scale as large as 100~TeV. Finally, we illustrate by example models how ultraviolet completion could eventually generate the mentioned dim-6 operators.

\vspace{0.5cm}
\noindent
%%%%%%%%%%%%%%%%%%%%%%%
\section*{Acknowledgement}
%%%%%%%%%%%%%%%%%%%%%%%
This work was supported in part by the Grants No.~NSFC-11975130, No.~NSFC-11575089, by The National Key Research and Development Program of China under Grant No. 2017YFA0402200, by the CAS Center for Excellence in Particle Physics (CCEPP). Xiao-Dong Ma is supported by the MOST (Grant No.~MOST~106-2112-M-002-003-MY3).

%%%%%

\end{document}